\documentclass[pra,superscriptaddress,twocolumn,amsmath,amssymb,11pt,reprint]{revtex4-2}

\usepackage{amsmath}
\usepackage{mathrsfs}
\usepackage{bm}
\usepackage{hyperref}
\usepackage{graphicx}
\usepackage{natbib}
\usepackage{physics}
\usepackage{braket}
\usepackage{IEEEtrantools}

\usepackage[scr=boondoxo]{mathalpha}

\usepackage[dvipsnames]{xcolor}

\newcommand{\be}{\begin{equation}}
\newcommand{\ee}{\end{equation}}
\newcommand{\bea}{\begin{eqnarray}}
\newcommand{\eea}{\end{eqnarray}}
\newcommand{\ra}{\rangle}
\newcommand{\la}{\langle}
\newcommand{\ud}{\textrm{d}}
\newcommand{\Ai}{\textrm{Ai}}

\begin{document}

\author{Marko M. \'{C}osi\'{c}}
\affiliation{Vin\v ca Institute of Nuclear Sciences,  P.O. Box 522 11001, Belgrade, Serbia}
\author{Andrew N. Jordan}
\affiliation{Institute for Quantum Studies, Chapman University, Orange, California 92866, USA}
\affiliation{The Kennedy Chair in Physics, Chapman University, Orange, California 92866, USA}

\title{Caustics and Superenergy in the Quantum Bouncer}

\begin{abstract}\noindent
We investigate the quantum interference and energetic phenomena associated with classical caustics in the quantum bouncing ball problem, we refer to as quantum caustics.  By considering an initial Gaussian wavepacket, we show that caustics associated with the underlying classical trajectory families are exhibited.  We connect the associated phase singularity chains in the vicinity of the caustic with the semiclassical Pearcey function built on the cusp catastrophe lines.  We also quantify the amount of superenergy exhibited in these solutions - regions of space where the local energy exceeds the largest constituent energy eigenvalue.  We give a complimentary description of the caustic and superenergy behavior using the Madelung/Bohm trajectories, which gives additional insight about the energy of the trajectories and how they traverse the phase singularity chains.
\end{abstract}
\maketitle

\section{Introduction}\noindent
Caustic patterns in nature have fascinated people for millennia.  This effect underlies everyday phenomena ranging from rainbows to bright and dark pattens of light from a drinking glass.  Originating from the Greek word $\kappa\alpha\upsilon\sigma\tau\acute{o}\varsigma$, meaning ``to burn'', the word evokes the convergence of sunlight rays of a magnifying glass or concave mirror to ignite a fire, a technology known at least since the ancient Greeks, notably described by Anthemius of Tralles in his treatise ``On Burning Mirrors'' \cite{knorr1983geometry}.  The caustic effect originates as an envelope of rays of light, or set of trajectories in mechanical systems, creating bright or dark regions, whose intensities can diverge. Underlying both of these examples is a wave theory, so it is natural to inquire how these caustics are diffraction softened in any natural phenomena that has a wavelike description. In this article, we focus on the field of quantum mechanics, and how ``quantum caustics'' naturally appear in the vicinity of the analogous classical caustic.  The structure of the classical caustic is closely connected to branching of paths in phase space \cite{arnold2013singularities}.  

With a fully quantum description, `quantum caustics' in monitored systems have previously been discussed in the description of the most likely path between initial and final quantum states, signaled by a catastrophe \cite{PostonStewart1996Catastrophe}, where the Lagrange manifold goes from being single valued to multivalued.  This effect in continuously monitored quantum systems has been described both theoretically and verified experimentally in Ref.~\cite{naghiloo2017quantum}.   It is possible that the catastrophes can multiply, resulting a bona fide chaotic quantum systems \cite{lewalle2018chaos}.

For isolated quantum systems, delicate interference effects can arise in the vicinity of associated classical caustics, giving rise to chains of phase singularities and bright and dark structures that organize around classical cusp singularities \cite{berry1980iv}.  We will focus in this article on describing such effects in the quantum bouncer -- a model describing a particle confined by a linear potential and an impenetrable boundary \cite{Langhoff1971,Gibbs1975} -- as a better representative of typical bound‑state dynamics.  The quantum bouncer model has immediate applications to experiments with cold neutrons \cite{abele2009qubounce}, and is a prototypical example for implementing a quantum measurement engines as an elevator \cite{elouard2018efficient}.  The two dimensional version of the bouncing ball exhibits a paraboloidal caustic, a version of which we study here \cite{berry1982wavelength}.  A pedagogical presentation of the quantum bouncer problem is discussed in Ref.~\cite{gea1999quantum}, with a focus on decay and revival of oscillations of the expected position of the ball for a Gaussian initial wavepacket.
\par
To help guide the investigation as to the physical significance of these interference patterns, we will also consider to what extent these structure exhibit non-classical features.  We quantify this with the concept of {\it super-phenomena} of observables \cite{Superphenomena}.  This is defined as when the local values of an observable (that will be defined rigorously below) exceeds the eigenvalue bounds of the quantum state, closely related to weak values \cite{Aharonov1988PRL,Dressel2014Colloquium,jordan2024quantum}.  In this article, we will focus on the energy of the system.

The article is organized as follows:  In Sec.~\ref{sec:lobserv} we define the concept of local observables and what we mean by ``superbehavior''.  We give examples of superoscillations - corresponding to superbehaving momentum, and also superbehavior in energy.  In Sec.~\ref{sec:bounce}, we present and solve the energy eigenvalue problem for the quantum bouncing ball, and also present its classical solution as well.  In Sec.~\ref{sec:qcaustics} the initial state of the bouncer is taken to be a Gaussian wavepacket, and we show how the quantum caustics naturally appear, and how they are closely connected with the branching structure of the classical catastrophe lines. Local energy behavior is also discussed.  In Sec.~\ref{sec:MBtraj} we analyze this problem also from a quantum hydrodynamic perspective, introduced by Madelung \cite{madelung1927quantum}, and refined into an interpretation by Bohm \cite{Bohm1952I,Bohm1952II}.  This trajectory (or streamline) based viewpoint brings new insights into the physics.  We conclude in Sec.~\ref{sec:conc}.

\section{Local Observables and Superphenomena} \label{sec:lobserv}\noindent
We now make precise the notions of the introduction by defining the local value $\tilde O$ of the observable $\hat O$ with discrete spectrum $\{o_1, o_2, \ldots\}$ at position $x$ for quantum state $|\psi\ra=\sum_nc_n|o_n\ra$ is defined as weak value of $\hat O$ with postselected position \cite{Aharonov1988PRL,Dressel2014Colloquium,Superphenomena,jordan2024quantum}
\begin{equation}\label{Eq-WV}
    \tilde O(x)=\frac{\la x|\hat O|\psi\ra}{\la x|\psi\ra}=\frac{\sum_nc_no_n\la x|o_n\ra}{\sum_nc_n\la x|o_n\ra}.
\end{equation}
Here $\{c_n\}$ are arbitrary complex coefficients, such that the state is normalized. The case of a continuous spectrum is defined in an analogous way by the replacement of the sum by an integral.
In the context of weak measurement $\Re\{\tilde O(x)\}$ determines the shift of the pointer wave function \cite{Dressel2014Colloquium}, while $\Im\{\tilde O(x)\}$ quantifies the change of shape of the pointer wave function induced by the measurement back-action \cite{Dressel2012ImaginaryWV}.
\par
Although the sum in Eq.~(\ref{Eq-WV}) formally extends over the entire spectrum, the reliable control of states $|o_n\ra$ for $n\gg 1$ is often impossible. 
We therefore restrict attention to $|\psi\ra$ belonging to a finite subspace spanned by vectors $|o_\textrm{min}\ra$, \ldots, $|o_\textrm{max}\ra$.
Note, however, that both $\Re\{\tilde O(x)\}$ and $\Im\{\tilde O(x)\}$ can exceed the spectral range $[o_\textrm{min},o_\textrm{max}]$, as famously demonstrated in spin weak-measurement \emph{gedankenexperiment} \cite{Aharonov1988PRL}. 
Whenever this occurs, we say that the state $|\psi\ra$ superbehaves at $x$ with respect to $\hat O$, and refer to the corresponding local values $\tilde O(x)$ as a superobservable \cite{Superphenomena}.
Although superbehavior can arise even in superpositions of two states \cite{Superphenomena}, the most interesting regime emerges in the limit when the finite spectral range $[o_\textrm{min},o_\textrm{max}]$ becomes populated with an increasingly large number of states.
Let $\hat O_1$ and $\hat O_2$ be two arbitrary observables.
If $\hat O_1\hat O_2-\hat O_2\hat O_1\neq0$ then superbehavior of $|\psi\ra$ in the respect to $\hat O_1$ does not imply superbehavior in the respect to $\hat O_2$.
However, according to the definition (\ref{Eq-WV}), both observables must simultaneously superbehave in the vicinity of all wave function nodes, that is, near points where $\la x|\psi\ra=\psi(x,t)=0$, unless the numerator of (\ref{Eq-WV}) also vanishes.
\par
To illustrate these concepts, consider the operator $\hat E=i\hbar\partial_t$, where $\hbar$ is the reduced Planck constant and $t$ is time. We further consider a wavefunction written in polar form $\psi(x,t)=R(x,t)\exp[iS(x,t)/\hbar]$.
The corresponding local energy function is given by
\begin{equation} \label{localE}
    \tilde E(x,t)=-\partial_t S(x,t)+i\hbar\partial_t\log R(x,t).
\end{equation}
Since quantum dynamics is inherently random, the energy of particles detected at the spacetime point $(x,t)$ is a random variable that fluctuates around the mean value.
The real part of the weak value $\Re\{\tilde E(x,t)\}$ is controlled by the phase and may be interpreted as the local mean energy conditioned on finding the particle at position $x$.  This interpretation takes on operational significance if a weak measurement of the particle's energy is made, with a subsequent post-selection of the particle as position $x$:  The pointer's conditioned average corresponds to the real part of the local energy \cite{jordan2024quantum,Dressel2014Colloquium}. 
The imaginary part, $\Im\{\tilde E(x,t)\}$, has no direct classical counterpart.
It can be interpreted as energy associated with the redistribution of particles reflected in the time-dependent changes in the shape of the probability density $\rho(x,t)=R^2(x,t)$ \cite{Dressel2012ImaginaryWV}. 
Superbehavior of $\Re\{\tilde E(x,t)\}$ has been interpreted as a nonzero probability of detecting a gamma photon emerging out of the cavity consisting only of infrared photons from a slit at position $x$ \cite{Aharonov1990InfraredGama,Berry2018EscSuperOsc,Aharonov2021Conservation}.
By contrast, the superbehavior of $\Im\{\tilde E(x,t)\}$ manifests as anomalously rapid growth or decay of the wavefunction amplitude, a phenomenon commonly referred to as supergrowth or superdecay, respectively \cite{Jordan2020Superresolution,Karmakar2023Supergrowth}.  Superoscillations have applications in optical superresolution - for a recent review, see Ref.~\cite{JordanHowellVamivakasKarimi2025}.
\par
The simplest way to create a superbehaving region in space is to engineer a state with a prescribed distribution of nodes. 
For example, a $n$-order zero of the Bessel wave $\psi(r,\varphi)=J_n(r)\exp[in\varphi]$ at $r=0$, (with $r$ and $\varphi$ polar coordinates) when perturbed by $\kappa J_0(r)$, with $0<\kappa\ll 1$, splits into $n$ simple zeros uniformly arranged on a circle of radius $r=2(\kappa n!)^{1/n}$ that by can be made arbitrarily small \cite{Berry2013BesselSO}.
\par
In a more involved approach, superbehaving functions are constructed from functions with band-limited Fourier spectrum. 
The canonical example of such a function with a discrete Fourier spectrum is 
\begin{equation}
    f_N(x)=\left(\cos(\frac{x}N)+ia\sin(\frac{x}{N})\right)^N\xrightarrow[N\gg1]{}\exp[iax],
\end{equation}
which superbehaves for $a>1$ with respect to the operator $\hat k_x=-i\partial_x$, within an arbitrarily large finite region around the point $x=0$, despite its spectrum being strictly limited to $-1\leq o_n\leq 1$ \cite{Aharonov1988PRL}. 
This function appears naturally for a particle confined to move in a ring with respect to the operator $\hat L/\hbar=-i\partial_\varphi$. 
In this case, the spectrum is already discrete due to the periodic boundary condition \cite{aharonov2023conservation}.
For this class of functions and operators, superphenomena are usually referred to as superoscillations (in space, angle, etc.).
\par
A more general example of a superbehaving function 
\begin{equation}\label{Eq-SuperEFun}
\begin{IEEEeqnarraybox}[][c]{rCl}
    g_N(x) &=&\exp\left[-\frac{x^2}{2N}\right]\left(1+ia\frac{x}{\sqrt{N}}\right)^N \nonumber\\
    &&\xrightarrow[N\gg 1]{}\exp\left[-\frac{x^2}{2N}+ia\sqrt{N}x\right],
\end{IEEEeqnarraybox}
\end{equation}
is constructed from the first $N$ eigenstates of the dimensionless harmonic oscillator Hamiltonian, $\hat H=\frac{1}{2}(-\partial_x^2+ x^2/N)$, exhibiting the local energy $\tilde E(x)\approx a^2 N /2$ close to the point $x=0$ \cite{Superphenomena}.  Although the maximum energy in the constituent sum is constant with respect to $N$, the local energy near the origin grows linearly with $N$.
Strictly speaking, the function (\ref{Eq-SuperEFun}) is not band-limited with respect to $x$, although Gauss-regularized functions are often considered in engineering practice as approximations of band-limited functions \cite{Sanner1992GaussControl}.  It is however, {\it energy-limited}, meaning that it can be constructed with energy eigenfunctions bounded in their energy eigenvalues.
However, the evolution of (\ref{Eq-SuperEFun}) is band-limited with respect to $t$ - which is equivalent to being energy-limited.
The time-dependent Schr\"odinger equation $i\hbar \partial_t |\psi\ra=\hat H|\psi\ra$ holds for all states, so superbehavior of energy is always accompanied with superoscillation in time.
Similar functions can be constructed using spectrally limited superpositions of eigenfunctions of the total orbital momentum operator $\hat L^2/\hbar^2=-\partial^2_{\theta}-\cot\theta\partial_\theta-\csc^2\theta\partial^2_{\varphi}$, whose spectrum is quadratic $o_\ell = \ell (\ell+1)$, with degeneracy $2\ell +1$. Here, $\ell = 0, 1, 2, \ldots$ and $\theta,\phi$ denote polar and azimuthal angles, respectively \cite{Superphenomena}. 
\par
Continuous band-limited superpositions, such as those creating random optical speckle patterns, are found to be superoscillatory in the regions containing phase singularities, which occupy approximately 1/3 of the two-dimensional wave field \cite{Dennis2008SuperoscillationSpeckle,Berry2020SuperoscillationsStanding}. Similarly, supergrowth was experimentally shown to also occur in speckle patterns \cite{viteri2024supergrowth}.
It was shown that an arbitrary superoscillatory or supergrowing function -- whether local or global -- can be constructed as a superposition of spherical Bessel functions $j_n(x)$ with a continuous but band-limited Fourier spectrum \cite{Karmakar2023Beyond}. 
These functions are eigensolutions of the radial Schr\"odinger equation for a free particle in three-dimensional space \cite{SakuraiNapolitano2011}.
It should be noted that there are many methods for constructing superbehaving functions based on the interpolation techniques (see, for example, Ref.~\cite{SodaKempf2020,JordanHowellVamivakasKarimi2025}) that are not directly related to eigenfunctions of operators commonly appearing in quantum mechanics and will not be discussed further.
\par
Our goal here is to examine the superbehavior of a generic bound quantum system with a discrete spectrum.
Note that none of the previously mentioned studies are applicable to the stated problem for two major reasons.
Firstly, the superbehavior observed in free-particle dynamics does not readily generalize to bound systems, because the boundary conditions of free and bound states are fundamentally incompatible.
Additionally, inclusion of the confining potential to the system Hamiltonian produces wave functions $|\psi\ra$ that are not necessarily band-limited.
Secondly, the dynamics of the harmonic oscillator is isochronic, meaning that all initial states evolve with the same period.  
Such systems are exceptional because superintegrability in classical systems requires the existence of additional globally defined integrals of motion, a property satisfied by only a small number of systems \cite{Gonera2004SuperInt,GoneraKosinskiMaslanka2001}.
In the quantum case, superintegrability requires the existence of additional globally defined symmetry operators compatible with the Hamiltonian, a highly restrictive condition that only a few models satisfy \cite{SheftelTempestaWinternitz2001}.
\par
The spectrum of the bouncer’s Hamiltonian is discrete and grows asymptotically as $E_n \sim n^{2/3}$, which generically results in aperiodic dynamics. 
Its eigenfunctions, expressed in terms of Airy functions \cite{vallee2010airy}, are not band‑limited, thereby precluding an investigation of superoscillatory behavior.
Instead, we focus on the superbehavior of the continuous spectrum of local energy $\tilde{E}(x,t)$ relative to the prepared state Eq.~(\ref{localE}).
As will be shown later, the evolution of the Gaussian wave packet produces an intricate distribution of phase singularities, in the vicinity of which the local energy spectrum $\tilde{E}(x,t)$ exhibits superbehavior.
It will be shown that the form of the energy eigenfunction is sufficiently simple to permit an analytical analysis of the conditions under which spectrally limited superposition exhibits superbehavior in a extended region of space and time.

\section{Quantum Bouncer} \label{sec:bounce}

\noindent
In this section, we formulate and solve the basic classical and quantum physics of the quantum bouncer problem.
Let $\psi(x,0)=\psi_0(x)$ be an initial state of non-relativistic quantum particle of mass $m$, confined to move in the half-line by the external potential
\begin{equation}\label{Eq-Pot}
    V(x) = \left\{\begin{IEEEeqnarraybox}[][c]{c,s,c}
       m g x,&\quad for &x\ge0,\\
        \infty,&\quad for&x<0,\end{IEEEeqnarraybox}\right.,
\end{equation}
where $-g$ is the gravitational acceleration.
Its subsequent evolution in time satisfies Schr\"odinger equation
\begin{equation}\label{Eq-SchEq}
    i\hbar\partial_t\psi(x,t)=-\frac{\hbar^2}{2m}\partial_x^2\psi(x,t)+mgx\cdot\psi(x,t),
\end{equation}
with the boundary condition $\psi(0,t)=0$. 
When $N$ is sufficiently large, an arbitrary solution of Eq. (\ref{Eq-SchEq}) can be accurately represented by a finite superposition \cite{SakuraiNapolitano2011}
\begin{equation}\label{Eq-Psi_x_t}
    \psi(\chi,\tau)\approx\psi_N(x,t)=\sum_{n=1}^N c_n\varphi_n(x)\exp\left[-\frac{i}{\hbar}E_nt\right],
\end{equation}
where $\varphi_n(x)$ and $E_n$ are solutions of the following eigenvalue problem
\begin{equation}\label{Eq-4}
    -\frac{\hbar^2}{2m}\partial_x^2\varphi_n(x,t)+mgx\cdot\varphi_n(x,t)=E_n\varphi_n(x,t),
\end{equation}
and 
\be\label{Eq-C_n}
c_n = \int\limits_{0}^\infty\varphi_n^*(x)\psi_0(x)\ud x.
\ee
For convenience, we introduce the dimensionless variables $\chi=x/L$, $\tau=t/T$ and $\varepsilon_n=E_n/E$ with characteristic units of length, time and energy satisfying the following constraints 
\begin{equation}
    \frac{\hbar^2}{2mL^2E}=1,\quad TE=\hbar,
\end{equation}
and set $\alpha=mgL/E$.
To simplify the notation, we shall often use prime ${}'$ and over-dot $\dot{}$ to denote $\partial_\chi$ and $\partial_\tau$, respectively.
In these units, the system's Hamiltonian is given by
\begin{equation}
    \hat H = - \partial_\chi^2 + \alpha\chi,
\label{ham}
\end{equation}
and the eigenvalue problem (\ref{Eq-4}) now reduces to a variant of the Airy equation \cite{vallee2010airy}
\begin{equation}
    \varphi''_n(\chi)-\alpha\left(\chi-\frac{\varepsilon_n}{\alpha}\right)\varphi_n(\chi)=0,
\end{equation}
whose solution is 
\begin{IEEEeqnarray}{rCl}\label{Eq-phi_n}
        \varphi_n(\chi) &=& \frac{\alpha^{1/6}}{|\Ai'(z_n)|}\Ai(\alpha^{1/3}\chi+z_n),\\
        z_n&\approx&-[3\pi/2(n-1/4)]^{2/3},    
\end{IEEEeqnarray}
where $\Ai(z_n)=0$, and $\varepsilon_n=-\alpha^{2/3}z_n$.
Note that Airy zeros $z_n$ are negative, which consequently yield positive values of $\varepsilon_n$, as expected.
\par
The trajectory of the classical bouncer is governed by Newton's equation of motion
\begin{equation}
    \ddot{\chi}+2\alpha=0,
\end{equation}
and the initial conditions $\chi(0)=\chi_0$ and $\dot\chi(0)=\dot\chi_0$, whose solution is given by the periodic parabolic function
\begin{IEEEeqnarray}[]{c}\label{Eq-x(t)}
    \chi(\tau;\chi_0,\dot\chi_0) = \textrm{pb}(\tau-\delta\tau),\\
    \textrm{pb}(\tau)=\begin{cases}
        \chi_\textrm{m}-\alpha\left(\tau-T/2\right)^2,&\quad\tau\leq T,\\
        \textrm{pb}(\tau-T),&\quad\tau>T,
    \end{cases}.
\end{IEEEeqnarray}
where $\delta\tau=(\dot\chi_\textrm{m}-\dot\chi_0)/2\alpha$ is the time delay, $T=\dot\chi_\textrm{m}/\alpha$ is trajectory period, while $\chi_\textrm{m}$ and $\dot\chi_\textrm{m}$ are maximal height and maxima intensity of velocity, related by the conservation of energy
\begin{equation}
    \dot\chi_0^2+4\alpha\chi_0=4\alpha\chi_\textrm{m}=\dot\chi_\textrm{m}^2.
\end{equation}
\par

\section{Quantum caustics} \label{sec:qcaustics}
\noindent
\par\noindent
In this section, we investigate the distribution of phase singularities generated by the evolution of a bouncing ball represented initially by the following Gaussian wave packet
\begin{equation}\label{Eq-Psi_0}
    \psi_0(\chi) = \frac{1}{\sqrt[4]{2\pi\sigma^2}}\exp\left[-\frac{(\chi-\chi_0)^2}{4\sigma^2}\right],
\end{equation}
of mean position $\chi_0$, and small variance $\sigma^2\ll 1$.
In that case, the lower limit of the integral in Eq. (\ref{Eq-C_n}) can be extended to $-\infty$, and the resulting integrals evaluated analytically, giving
\begin{widetext}
\begin{align}
\label{Eq-Gauss_Cn}
    c_n\approx\frac{\sqrt[4]{8\pi\sigma^2}}{|\Ai'(z_n)|}\exp\left[\frac{2}{3}\sigma^6 \alpha^2 +(\alpha^{1/3}\chi_0+z_n)\alpha^{2/3}\sigma^2\right]\Ai\left(\alpha^{4/3} \sigma^4+\alpha^{1/3} \chi_0+z_n\right).
\end{align}
\end{widetext}

The details of derivation are given in appendix \ref{App-AF_Gauss}.
\subsection{Superoscillating initial state}
\begin{figure*}[!t]
    \centering
    \includegraphics[width=\linewidth]{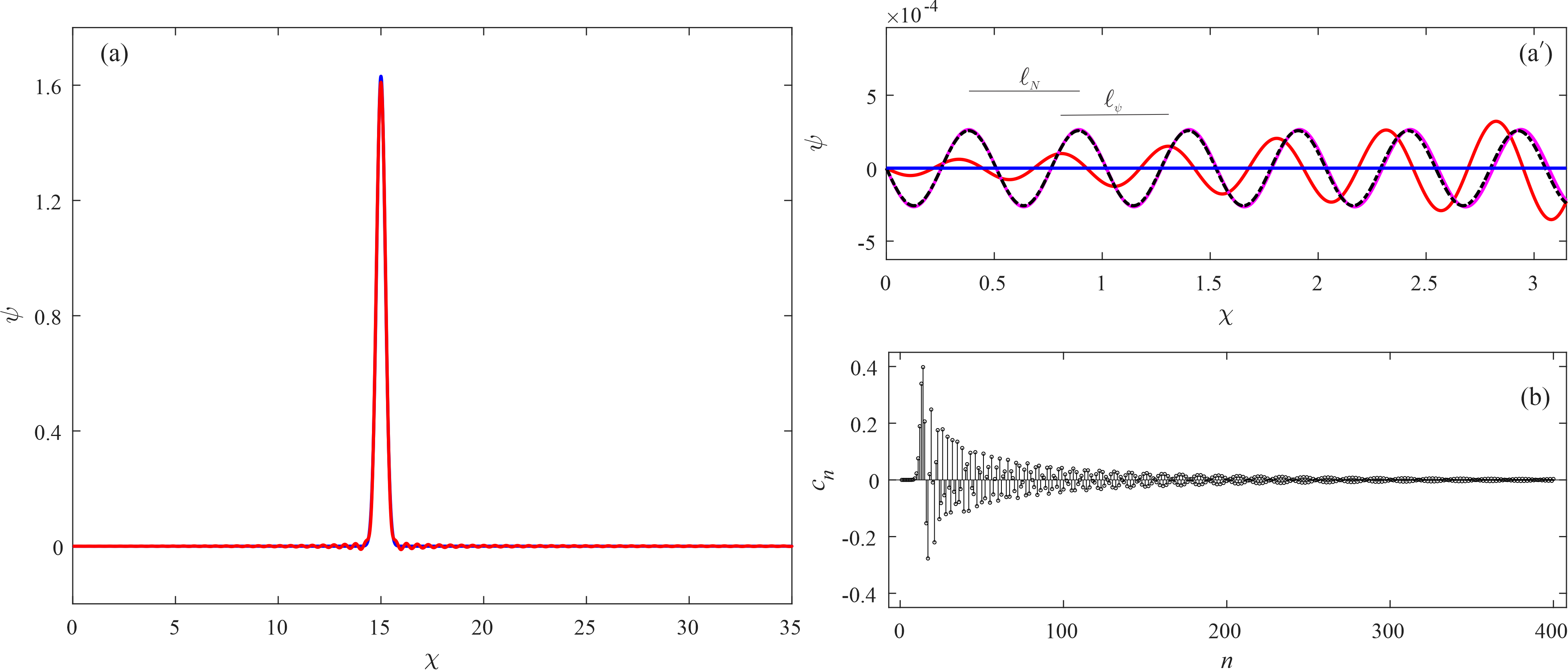}
    \caption{(a) The initial state $\psi_0(\tau)$ (the blue line), and its corresponding energy-limited approximation $\psi_N(\chi,0)$ (the red line) for $\chi_0=15$, $\sigma=0.15$, and $N=400$. 
    (a') Enlarged view in the vicinity of the coordinate origin. 
    The magenta line shows $c_{N}\varphi_{N}(\chi)$, its harmonic approximation $A_N\sin(|z_N|\chi)$ is shown by the dashed black line.
    (b) Stem plot of expansion coefficients $c_n$ calculated according to Eq. (\ref{Eq-Gauss_Cn}).    
    Lines $\ell_N$ and $\ell_\psi$ indicate the local wavelengths of $c_{N}\varphi_{N}(\chi)$ and $\psi_N(\chi,0)$ respectively.
    }
    \label{Fig-Initial_State}
\end{figure*}

\noindent
In this subsection, we set $\alpha=1$ for simplicity.
Figure \ref{Fig-Initial_State}(a), shows the initial state $\psi_0(\chi)$ and its approximation $\psi_N(\chi,0)$ for $\chi_0=15$, $\sigma=0.15$, and $N=400$.
The corresponding expansion coefficients $c_n$ are shown in Fig. \ref{Fig-Initial_State}(b).
For $n<9$, only the exponentially decaying side of $\varphi_n(\chi)$ overlaps with the center of the Gaussian $\psi_0(\chi)$, thus the corresponding $c_n$ are negligible. 
We can conclude that also from the formula (\ref{Eq-C_n}) because for $\eta=\sigma^4+\chi_0+z_n\gg0$, the super-exponential decay of $\Ai(\eta)\sim\exp[-\frac{3}{3}\eta^{3/2}]$ makes $|c_n|\approx 0$ \cite{vallee2010airy}.
The large values of $c_n$, as indicated by (\ref{Eq-C_n}), are expected for $\eta\approx0$.
These occur for $n=9,\ldots,16$, with the maximal amplitude $c_{14}=0.398$ belonging to eigenstate $\varphi_{14}(\chi)$ having principal maximum -- determined by the first zero of $\Ai'$ at $z_1'=-1.01879\:29716$ -- located at $z_1'-z_{14}=15.114$, which is remarkably close to $\chi_0$.
For moderate values of $n$, and $\eta\sigma^2\approx0$, $c_n$ is an oscillatory function decaying like $|c_n|<1/\sqrt{\eta}$, that can be deduced from the asymptotic formulas for $\Ai(\eta)$ and $\Ai'(\eta)$ when $\eta$ is large and negative \cite{vallee2010airy}.
Lastly for $n\gg1$, $\eta\lll-1$, the decay rate accelerates to $|c_n|\sim\exp[\sigma^2z_n]$.
Thus, for $\sigma$ small, a large number of eigenvalues are necessary to accurately represent the initial state (\ref{Eq-Psi_0}).
The norm of the approximation $|\psi_N(\chi)|^2=\sum_{n=1}^Nc_n^2=0.9996$ is very close to the ideal value 1.
Additional numerical experiments showed that the error of the norm decreases with $N>400$ and that the approximation (\ref{Eq-Gauss_Cn}) gives reasonably accurate results even in the extreme case when $\chi_0\approx\sigma$.
\par
\par
Although the expression (\ref{Eq-Gauss_Cn}) provides an excellent approximation of the integral (\ref{Eq-C_n}), the oscillatory nature of the Airy function prevents $\psi_N(\chi)$ from converging monotonically to the limiting function $\psi_0(\chi)$.
Overshoots and undershoots of $\psi_N(\chi)$ -- visible in Fig.~\ref{Fig-Initial_State}(a) only near  $\chi_0$ -- actually extend across the entire domain, as seen from the magnified view of $\psi_N(\chi)$ and $\psi_0(\chi)$ near the origin in Fig.~\ref{Fig-Initial_State}(a$'$).
The oscillatory convergence of the finite approximations is commonly associated with the Gibbs or Runge phenomena \cite{gottlieb1977numerical,cheney2000course}.
We shall show now that it can also give rise to superoscillations.
Let us focus on the region $0\leq\chi\leq -z_1$ where $\chi$ can be considered small. 
As shown in appendix \ref{App-small_chi_asympt}, asymptotic formulas for $\Ai$ and $\Ai'$, allow $\psi_N(\chi)$ to be rewritten as a generalized Fourier sequence \cite{Aharonov2017Mathematics} 
\begin{equation}\label{Eq-Fur_Sec-approx}
\psi_N(\chi)\approx \sum_{n=1}^NA_n\sin\left(|z_n|^{1/2}\chi\right),
\end{equation}
with the expansion coefficients 
\begin{equation}\label{Eq-Exp_coeff}
    A_n=(8\pi^3\sigma^2)^{1/4} \frac{(-1)^{n+1}}{|z_n|^{3/4}}e^{(\chi_0 +z_n)\sigma^2}\mathrm{Ai}(\chi_0 + z_n).
\end{equation}
Despite the crude approximations used in deriving Eq.~(\ref{Eq-Fur_Sec-approx}), Fig.~\ref{Fig-Initial_State} (a$'$) shows it remains valid well beyond the originally intended range.
Using the harmonic addition theorem
\begin{equation}
    \sum_{n=1}^N A_n\sin(\eta+b_n)= A\sin(\eta+b),
\end{equation}
where 
\begin{IEEEeqnarray}{c}
A^2=\sum_{n=1}^N\sum_{m=1}^NA_nA_m\cos(b_m-b_n),\\
\tan b=\frac{\sum_{n=1}^NA_n\sin b_n}{\sum_{n=1}^NA_n\cos b_n},
\end{IEEEeqnarray}
sum (\ref{Eq-Fur_Sec-approx}) can be expressed formally using a single harmonic function
\begin{equation}\label{Eq-harm_approx}
    \psi_N(\chi) \approx A_\psi\sin\phi_\psi(\chi),
\end{equation}
whose amplitude and phase functions are given by the relations
\begin{IEEEeqnarray}{c}
    A_\psi^2=\sum_{n=1}^N\sum_{m=1}^NA_nA_m\cos((|z_m|^{1/2}-|z_n|^{1/2})\chi),\\ \tan\phi_\psi(\chi)=\frac{\sum_{n=1}^NA_n\sin (|z_n|^{1/2}\chi)}{\sum_{n=1}^NA_n\cos (|z_n|^{1/2}\chi)},
\end{IEEEeqnarray}
Note that because of continuity, it is possible to find $\chi$ such that $\phi(\chi)\approx k_\psi\chi$, for which the approximation  (\ref{Eq-harm_approx}) reduces to a simple trigonometric function.  Thus, we interpret $k_\psi$ as the wavenumber of the approximate wavefunction (\ref{Eq-harm_approx}). 
The simplest way to find $k_\psi$ is to note that it must be equal to the coefficient in front of the linear term in the $\chi\lll1$ asymptotics of $\phi_\psi(\chi)$, which can be found easily using the small-angle approximation of $\sin\eta\approx\eta$ and $\cos\eta\approx 1$, giving
\begin{equation}\label{Eq-loc_k}
    k_\psi \approx \frac{\sum_{n=1}^N|z_n|^{1/2}A_n}{\sum_{n=1}^NA_n}.
\end{equation}
\par
\begin{figure*}[t]
    \centering
    \includegraphics[width=\linewidth]{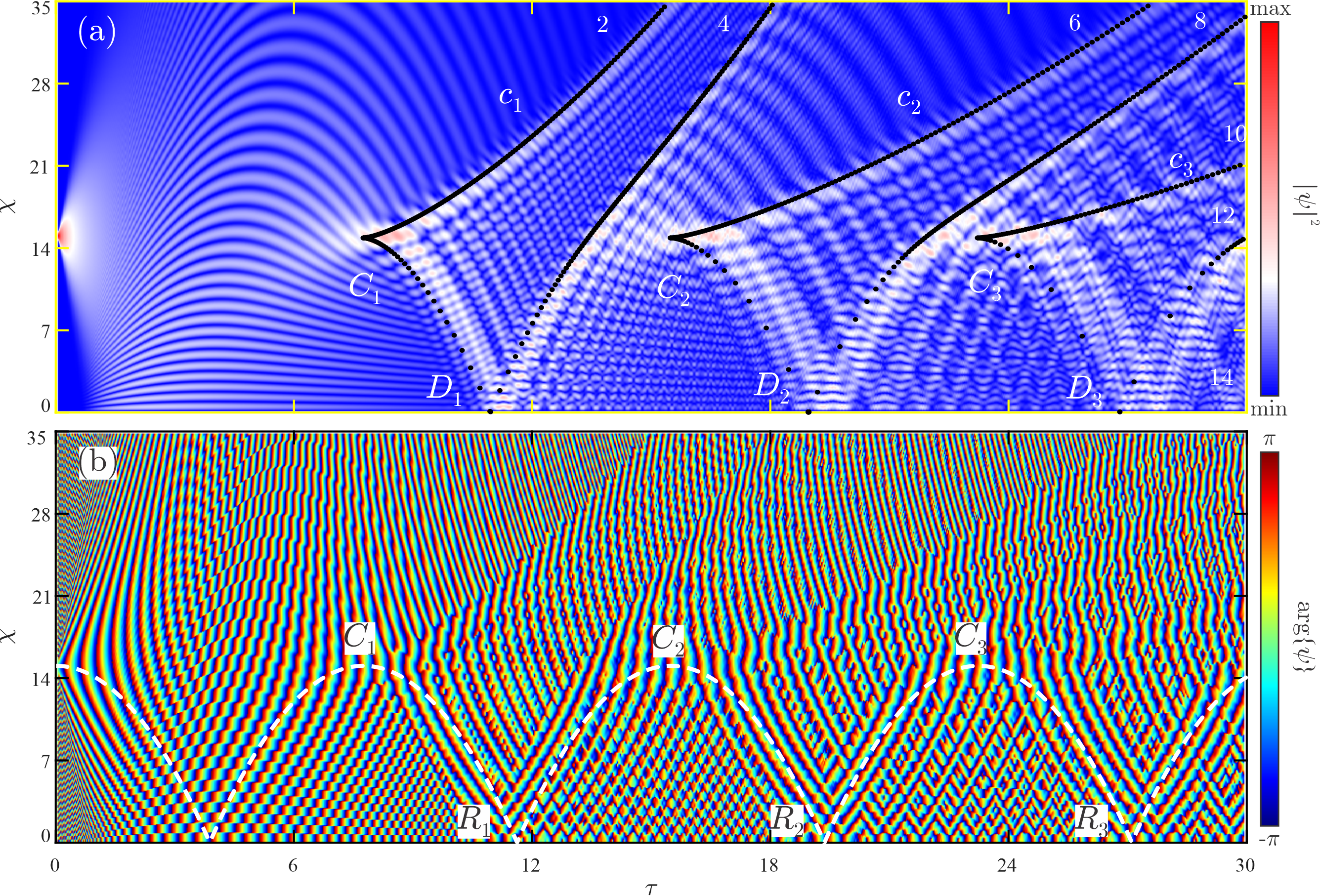}
    \caption{(a) Log-scale probability density $|\psi(\chi,\tau)|^2$ and (b) phase $\arg\{\psi(\chi,\tau)\}$ for $\chi_0=15$, $\sigma=0.15$, and $N=400$.
    Dotted black lines show the envelope of classical trajectories $\chi(\tau;\chi_0,\dot\chi_0)$ starting at $\chi_0$ with varying initial velocities. 
    Numbers denote point multiplicities in regions bounded by caustics $c_1, c_2, c_3$. The white dashed line is the reference trajectory $\chi(\tau;\chi_0,0)$.
    Cusps of caustics and turning points of the reference trajectory are labeled $C_1, C_2, C_3$. 
    Rebound points of downward caustic branches are $D_1, D_2, D_3$, with corresponding reference trajectory rebounds $R_1, R_2, R_3$.
    }
    \label{fig-Bouncing_state}
\end{figure*}
\par
The local wavelength of the highest excited state and the simple harmonic approximation (\ref{Eq-harm_approx}) are given by $\lambda_{N}=2\pi/|z_N|^{1/2}=0.509$ and $\lambda_\psi=2\pi/|k_\psi|=0.4317$, respectively, showing superoscillatory behavior. 
These wavenumber values are in excellent agreement with lengths of lines $|\ell_N|=0.504$, and $|\ell_\psi|=0.476$ in Fig. \ref{Fig-Initial_State} (a$'$), representing the distance between consecutive maxima of $c_N\varphi_N(\chi)$ and $\psi_N$, respectively.
Thus, the energy-limited approximation of $\psi_0$ superoscillates close to the coordinate origin whenever the denominator in Eq. (\ref{Eq-loc_k}) sums to a small value.
This is allowed because expansion coefficients (\ref{Eq-Exp_coeff}) are signed (they can be positive and negative).
However, our numerical experiments indicate $|k_\psi|\rightarrow|z_N|^{1/2}$ for $N\rightarrow\infty$, so no superoscillations exist in this limit.

\subsection{Quantum caustic pattern}\noindent
In this subsection, we give a quantitative description of the interference patterns of the quantum caustics, and connect it with classical catastrophe theory.
The figure \ref{fig-Bouncing_state} shows the evolution of $\psi_N(\chi,\tau)$ for $0\leq\tau\leq30$ and $0\leq\chi\leq35$, together with the classical trajectory of a particle starting at $\chi_0$ with zero initial velocity and period $T=2\sqrt{\chi_0}$.  Parameter values are the same as in the previous section; $N=400$ in particular.
The initial wavepacket expands rapidly, and for $1<\tau<8$, the overlap between the reflected part of the wave and the still-expanding component gives rise to the familiar interference fringes.
However, at the time of the first return point [see Fig. \ref{fig-Bouncing_state}(b)], the probability density spontaneously concentrates again at $\chi_0$ only to spit into two distinct \emph{streams} of probability current moving upward or downward, respectively. 
If these streams are interpreted as branches of the most probable trajectory after the return point, then the obtained solution suggests that the quantum particle becomes \emph{unstable} and splits into two parts. 
Since the upward branch is concave, a classical particle moving on it would experience an upward-directed effective force.
Similarly, the downward branches reach the impenetrable plane [points $D_1=(\tau_{D_1},\chi_{D_1})=(10.954,0)$, $D_2=(\tau_{D_2},\chi_{D_2})=(18.975,0)$, and $D_3=(\tau_{D_3},\chi_{D_3})=(26.835,0)$], faster than the reference classical trajectory [points $R_1=(\tau_{R_1},\chi_{R_1})=(3T/2=11.619,0)$, $R_2=(\tau_{R_2},\chi_{R_2})=(5T/2=19.365,0)$, and $R_3=(\tau_{R_3},\chi_{R_3})=(7T/2=27.111,0)$], and after the reflection, continue moving upward.
The maximal height reached by the streams is approximately $-z_N$, where they curve downwards, as confirmed by additional calculations performed with $N=600$ and $N=800$. 
This behavior is expected since for $\chi\gg-z_N$ the eigen-expansion (\ref{Eq-Psi_x_t}) transforms into
\begin{widetext}
\begin{align}
\label{Eq-Evanescent_waves}
    \psi(\chi,\tau)\approx \sum_{n=1}^N\frac{c_N}{|\Ai'(z_n)|}\frac{\chi^{-1/4}}{2\sqrt{\pi}}\exp[
-\frac{2}{3} \chi^{3/2}
- z_n \chi^{1/2}+iz_n\tau]\sim\frac{\exp[
-\frac{2}{3} \chi^{3/2}]}{\chi^{1/4}},
\end{align}
\end{widetext}
for which the probability current
{\color{red}}
\begin{equation}
    J(\chi,\tau)=2\Im\{\psi_N(\chi,\tau)^*\psi'_N(\chi,\tau)\},
\end{equation}
is negligible.
We therefore conclude that, in the limit $N\rightarrow\infty$, the described upward streams reaches infinite height.
The described evolution repeats qualitatively at all other classical rebounds.
\par
With such a large number of excited states, $\psi_N(\chi,\tau)$ is expected to exhibit semiclassical behavior \cite{BerryMount1972}.
To understand it better, we have treated $\chi_0$ as a fixed parameter and considered the map $\chi(\dot\chi_0;\tau,\chi_0)$ as a function of $\dot\chi_0$ that depends on the free parameter $\tau$.
For any $\tau$, the abscissas of its critical points $\dot\chi_0^{(r)}(\tau)$, $r=1,2,\ldots$, are solution of equation
\be\label{Eq-Env-1}
\left.\partial_{\dot\chi_0}\chi(\dot\chi_0;\tau,\chi_0)\right|_{\dot\chi_0=\dot\chi_0^{(r)}(\tau)}=0.
\ee
Geometrically, they define a set of curves in $(\tau,\dot\chi_0)$ space whose images in the observable $(\tau,\chi)$ space is set of curves
\begin{equation}\label{Eq-Env-2}
\!\!\dot\chi_0^{(r)}(\tau)\rightarrow \chi_r(\tau)=\chi\left(\tau;\chi_0,\dot\chi_0^{(r)}(\tau)\right),\;r=1,2,\ldots,
\end{equation}
that coincides with the envelope of a continuous one-parameter 1D function family \cite{BruceGiblin1992}.
In optics, envelopes are associated with line focusing phenomena that give rise to bright lines known as caustics.
Semiclassical caustics can be viewed as the diffraction-softened images of line foci in geometrical optics or, more generally, singularities of the classical differential cross-section \cite{Berry1976WavesThom,Berry1981SingularitiesWavesRays}.
\par
The obtained collection of the classical caustics is shown by the dotted black lines in Fig. \ref{fig-Bouncing_state}(a).
For $0<\tau<30$ there are three caustics labeled $c_1$, $c_2$, and $c_3$, each composed out of two branches -- one directed upwards and another initially pointed downwards that is reflected upward once it reaches $\chi=0$ boundary.
If no limit is placed on the value of $\dot\chi_0$, then all upward branches extend to infinity, which agrees with the similar estimate made for the upward streams of the probability density.
\par
Note that the following equation holds,
\begin{equation}
    \partial_{\dot\chi_0}\textrm{pb}(\tau-\delta\tau)=\frac{\dot\chi_0}{2}+\left(\tilde\tau-\dot\chi_\textrm{m}+\frac{\dot\chi_0}{2}\right)\!\!\left(2\frac{\dot\chi_0}{\dot\chi_\textrm{m}}-1\right),
\end{equation}
where $\tilde\tau=(\tau\:\textrm{mod}\:T)=\tau-\left\lfloor\frac{\tau}{T}\right\rfloor T.$
For the $\dot\chi_0=0$ trajectory, $\partial_{\dot\chi_0}\textrm{pb}(\tau-\delta\tau)=0$ for $\tau=nT$, ($n=1,2,\ldots$) at which $\chi(nT;\chi_0,0)=\chi_0$.
Consequently, points $C_1=(\tau_{C_1},\chi_{C_1})=(T,\chi_0)$, $C_2=(\tau_{C_2},\chi_{C_2})=(2T,\chi_0)$, \ldots, always belong to the envelope, and are precisely locations where the two branches of the caustic line meet. 
According to the theory, caustics partition space into regions with different numbers of trajectories passing through each space-time point \cite{BruceGiblin1992}.
As evident from Fig. \ref{fig-Bouncing_state}(a), passing over the caustic, the multiplicity of points changes abruptly by two, which causes the density of the trajectories to be infinite along the caustic lines.
The side of the caustic of the increased multiplicity is called the bright side of the caustic, while the other side is known as the dark side.
\par
Since the function family $\chi(\dot\chi_0;\tau;\chi_0)$ is continuous, the new cusp point appears via a saddle-node bifurcation when the saddle point becomes a degenerate critical point, after which it splits into a minimum-maximum pair.
According to the catastrophe theory, any structurally stable function family, depending on one independent variable $\eta$ (called state variable) and two parameters $x$ and $y$ (called control parameters), is locally equivalent to the following universal polynomial prototype \cite{Arnold1972IntegralsOR,PostonStewart1996Catastrophe}.  The singularity $A_3$ is the Arnol'd notation for the cusp catastrophe.  It is given by
\begin{equation}
    A_3(\eta;x,y) = \eta^4+x\eta^2+y\eta,
\end{equation}
and correctly describes the change in the number and type of critical points of the function family $\chi(\dot\chi_0;\tau;\chi_0)$. 
Application of the catastrophe theory requires embedding $\chi(\dot\chi_0;\tau;\chi_0)$ into the equilibrium set of catastrophe 
\be\label{Eq-A_3_equilib_set}
\frac{\ud}{\ud\eta}A_3(\eta;x,y)=0,
\ee
that is always possible because (\ref{Eq-A_3_equilib_set}) can be viewed as an implicit definition of a continuous family of functions depending on $\eta$ parametrized by $x$ and $y$.
Thus, catastrophic modeling amounts to specifying the set of functions $\eta(\dot\chi_0)$, $x(\chi,\tau)$, and $y(\chi,\tau)$ while set of points defined by the envelope conditions (\ref{Eq-Env-1}) and (\ref{Eq-Env-2}) coincide with the degenerate critical set of catastrophe $A_3(\eta)$, also known as a bifurcation set \cite{Arnold1972IntegralsOR,PostonStewart1996Catastrophe}
\begin{equation}\label{Eq-A_4_biff_set}
    \frac{\ud}{\ud\eta}A_3(\eta)=\frac{\ud^2}{\ud\eta^2}A_3(\eta)=0\Rightarrow 27y^2 + 8x^3=0.
\end{equation}
\par
\begin{figure*}[t]
    \centering
    \includegraphics[width=\linewidth]{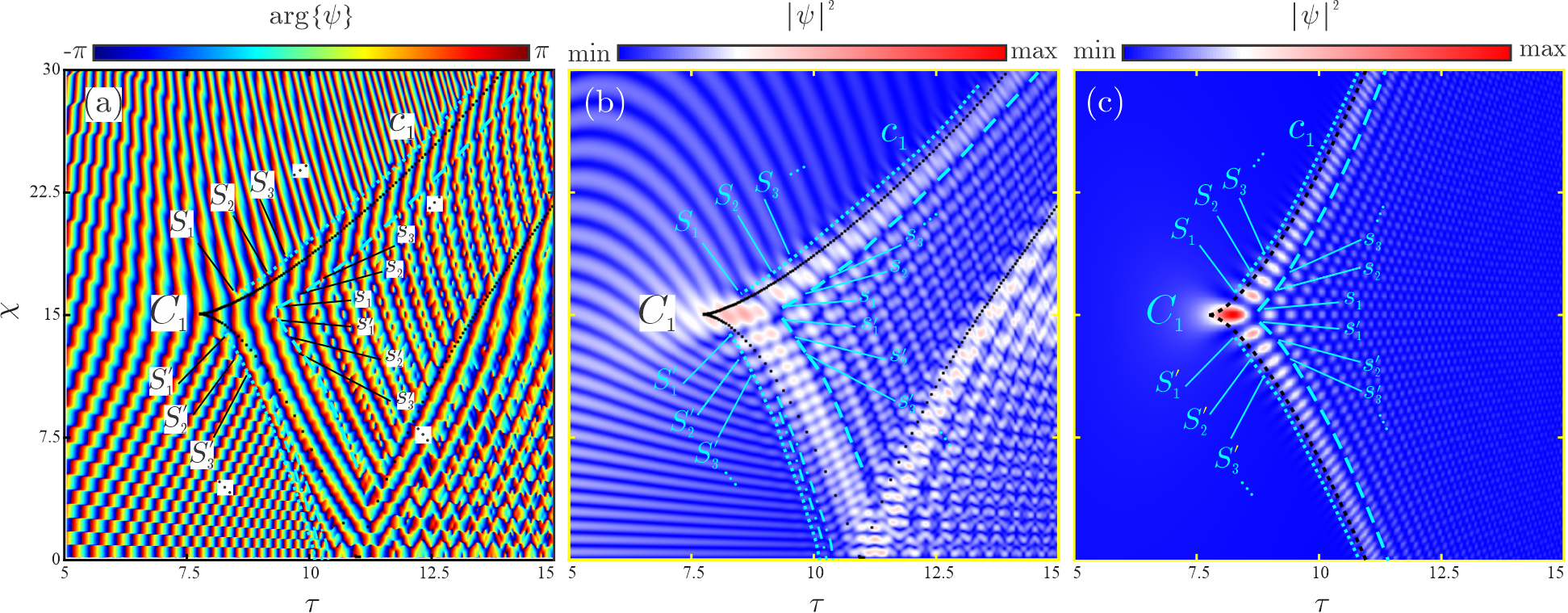}
    \caption{(a,b) Magnified views of $\textrm{arg}\{\psi_N(\chi,\tau)\}$ and  $|\psi_N(\chi,\tau)|^2$ near the cusp point $C_1=(\chi_0,2\sqrt{\chi_0})$.
    (c) The local catastrophic approximation $|\psi_N(\chi,\tau)|^2\approx|\textrm{P}((\tau_{C_1}-\tau)/\bar\tau_{C_1},(\chi-\chi_{C_1})/\bar\chi_{C_1})|^2$.
    Black dotted lines mark classical caustic lines in (a,b) and the bifurcation set of the optimal catastrophic polynomial $A_4$ in (c).
    Dashed and dotted cyan lines show singularity chains on the bright and dark sides of the caustics, respectively.
    Dot-dash cyan lines indicate additional singularity chains generated by four-wave interference.  We take $N=400$ in this plot.
    }
    \label{fig-Bouncing_state_zoom}
\end{figure*}
Equation (\ref{Eq-A_4_biff_set}) defines a curve in the $(x,y)$ space named semicubical (or sesquiplicate) parabola \cite{BruceGiblin1992}, which separate the parametric space into two regions where $A_3(\eta)$ has either one or three critical points, respectively. 
If functions $x(\chi,\tau)$ $y(\chi,\tau)$ can be chosen so that semicubical parabola (\ref{Eq-A_4_biff_set}) matches the caustic line, then an abrupt change in the number of critical points of $A_3$ will reflect the local change in the number of trajectories passing through a given point of spacetime $(\chi,\tau)$.
In general, constructing such a correspondence is difficult.
However, in the vicinity of a cusp point, the required functions take a particularly simple form $\eta\propto\dot\chi_0$, $x\propto\chi$ and $y\propto\tau$ (see appendix \ref{App-catatrophe-fit} for details).
According to the theory, the corresponding local catastrophic model of a semiclassical wave function is given by the oscillatory integral whose stationary points coincide with the critical set of the catastrophic polynomial \cite{Arnold1972IntegralsOR,PostonStewart1996Catastrophe}. 
Thus, for the cusp catastrophe, the appropriate local model of the semiclassical wavefunction is the Pearcey function \cite{pearcey1946structure,Berry2010Integrals}
\begin{IEEEeqnarray}{rCl}
    \textrm{P}(x,y)&=&\int\limits_{-\infty}^\infty\exp[iA_3(\eta)]\ud\eta
 \\   &=&\int\limits_{-\infty}^\infty\exp[i\eta^4+ix\eta^2+iy\eta]\ud\eta.
\end{IEEEeqnarray}

\par
Figures \ref{fig-Bouncing_state_zoom}(a) and (b) show enlarged views of $|\psi_N(\chi,\tau)|^2$ and $\textrm{arg}\{\psi_N(\chi,\tau)\}$ in the vicinity of the point $C_1$.
Note numerous points in Fig. \ref{fig-Bouncing_state_zoom}(a) where all equi-phase contours cross making $\textrm{arg}\{\psi_N(\chi,\tau)\}$ undefined.
These points occurs precisely where $|\psi_N(\chi,\tau)|^2=0$ and are known as wave nodes.
As shown in Figs. \ref{fig-Bouncing_state}(a) and Figs. \ref{fig-Bouncing_state_zoom}(a) and (b) the classical caustics form the skeleton of the quantum distribution $|\psi(\chi,\tau)|^2$, decorated by a regular distribution of wave nodes.
\par
On the bright side of the caustic, phase singularities appear in pairs, forming chains that run parallel to the caustic branches.
For clarity, Fig.~\ref{fig-Bouncing_state_zoom} highlights the first two such chains, formed by the singularities $s_1$, $s_2$, $s_3$, \ldots, and $s_1'$, $s_2'$, $s_3'$, \ldots, using dashed cyan lines.
Similar chains, formed by singularities $S_1$, $S_2$, $S_3$, \ldots, and $S_1'$, $S_2'$, $S_3'$, \ldots, respectively, appear on the dark side of the caustics and are highlighted by dotted cyan lines.
As expected, the phase singularities in Fig. \ref{fig-Bouncing_state_zoom}(a) correspond exactly to the wave nodes shown in Fig. \ref{fig-Bouncing_state_zoom}(b).
\par
It is shown in appendix \ref{App-catatrophe-fit}  that the simplest fit of the caustic line $c_1$ is given by semicubical parabola, adapted from Eq.~\ref{Eq-A_4_biff_set},
\begin{equation}
    27\left(\frac{\chi-\chi_{C_1}}{\bar\chi_{C_1}}\right)^2+8\left(\frac{\tau_{C_1}-\tau}{\bar\tau_{C_1}}\right)^3=0,
\end{equation}
with scale parameters $\bar\chi_{C_1}=0.509$ and $\bar\tau_{C_1}=0.224$. 
The corresponding probability density of the catastrophic model 
\begin{equation}
    \!\!\rho(\chi,\tau)=|\psi_N|^2\approx\left|\textrm{P}\left(\frac{\tau_{C_1}-\tau}{\bar\tau_{C_1}},\frac{\chi-\chi_{C_1}}{\bar\chi_{C_1}}\right)\right|^2\!,
\end{equation}
\par
\begin{figure*}[!t]
    \centering
    \includegraphics[width=1\linewidth]{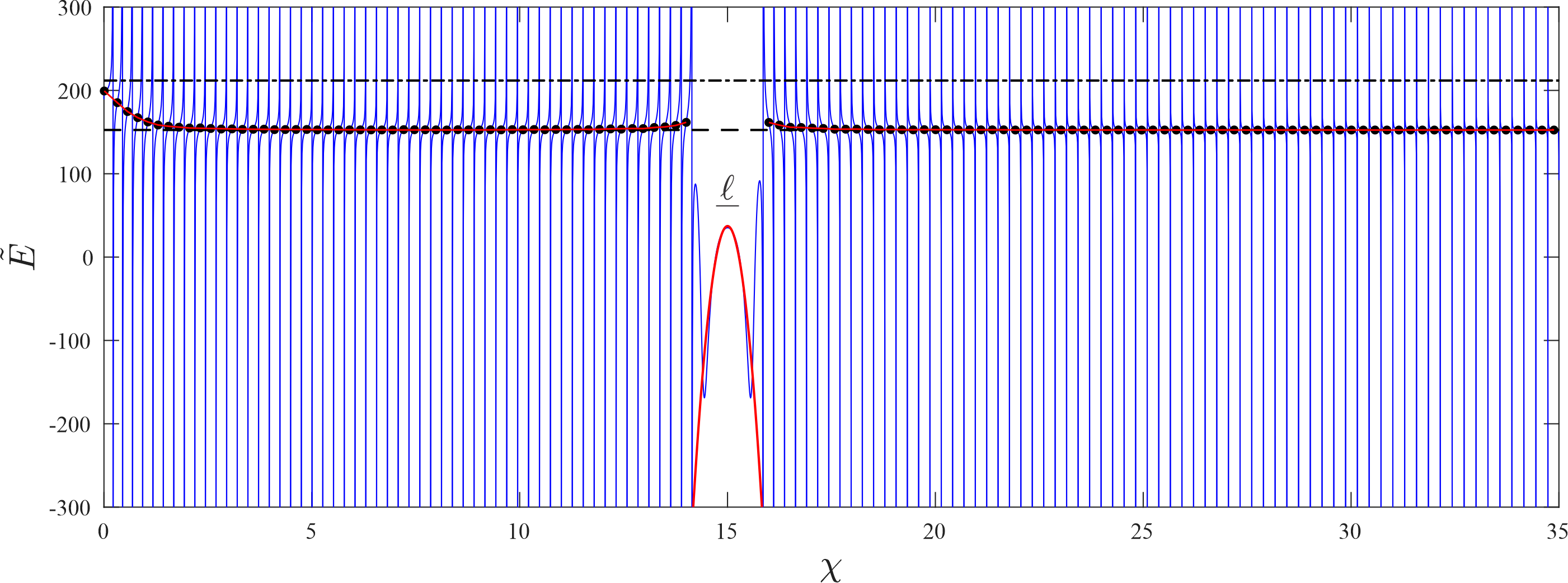}
    \caption{The local energy values $\tilde{E}_N(\chi,0)$ (blue) are plotted versus $\chi$, with its inflection point in the superbehaving region marked by black circles.
    The dot-dashed and dashed black lines indicate the energy levels $\varepsilon = k_\psi^2$ and $\varepsilon = \varepsilon_N$, respectively. 
    The limiting distribution $\tilde{E}(\chi,0)$ is shown in red.
    Line $\ell$ shows the span of the positivity interval (\ref{Eq-normal_Interval}).
    { The black-dotted-red line show the poly-lines $(\chi_{i_j},\tilde E(\chi_{i_j}))$.}
    }
    \label{fig-Init_Ew}
\end{figure*}
\par\noindent together with its bifurcation set are shown in Fig. \ref{fig-Bouncing_state_zoom}(c).
On the bright side of the caustics, its phase singularities are also organized in pairs and form chains running parallel with the branches of the bifurcation set.
A single singularity chain also appears on the dark side of the caustics.
Although the function $\textrm{P}(x,y)$ is modeled by three-real-wave interference inside the bifurcation set and by one real and two evanescent waves outside of it \cite{BerryNyeWright1979EllipticUmbilic}, it provides an excellent qualitative model for $\psi_N(\chi,\tau)$ close to the point $C_1$.
This is despite the fact that, according to Fig. \ref{fig-Bouncing_state}(a), the corresponding semiclassical model should involve four-wave interference.
As evident from Figs. \ref{fig-Bouncing_state_zoom}(a) and (b), close to $C_1$ the additional contribution of the reflection from the surface $\chi=0$ just stretches the distribution of singularities without creating any new ones.  
The deviation are noticeable only far from the point $C_1$ on the dark side of the lower caustic branch where 2-wave interference produces additional singularity chains [see the dot-dash cyan lines in Figs. \ref{fig-Bouncing_state_zoom}(a) and (b)].
\par
The periodic emergence of cusp points $C_1$, $C_2$, $C_3$, \ldots, is fully consistent with caustic evolution observed in other bound systems, such as the optical lattice potential \cite{Cosi2020PhaseSpaceCatastrophes}, as well as in the transmission of light through a sinusoidal volume grating \cite{BerryODell1999Ergodicity}.
The semiclassical model of the quantum caustics associated with points $C_2,\ldots$ can be constructed in exactly the same way as done for $C_1$. 
If necessary, the more elaborate models could be created by modeling the reflected caustics by a fold catastrophe $A_2$, and by modeling the global semiclassical wave as a superposition of local catastrophic models, each associated with a branch of the caustic line pattern \cite{Cosi2020PhaseSpaceCatastrophes}.
We shall not dwell any further on this issue since only the distribution of phase singularities will be important for our subsequent analysis of the superbehavior. 
\par
\begin{figure*}[!tb]
    \centering
    \includegraphics[width=\linewidth]{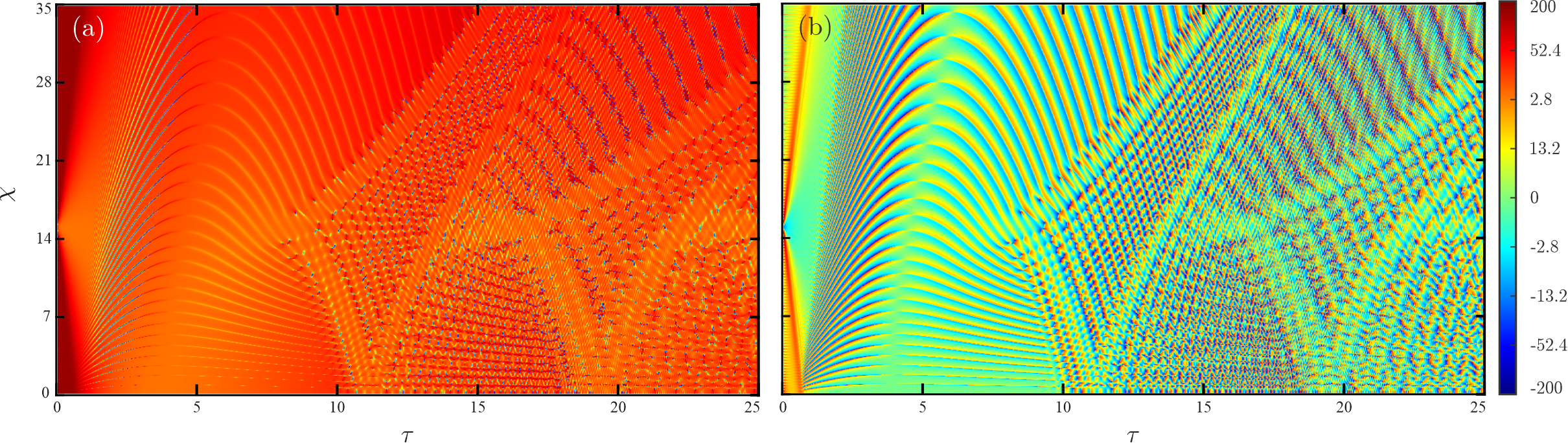}
    \caption{(a,b) Evolution of $\Re\{\tilde E_N(\chi,\tau)\}$ and $\Im\{\tilde E_N(\chi,\tau)\}$, respectively, for $\chi_0=15$, $\sigma=0.15$, and $N=400$. 
    For clearer visualization, both functions were restricted to the interval $[-200,200]$.  For this choice of parameters, $\varepsilon_N \approx 152.529$.
    }
    \label{Fig-Loc_Energy}
\end{figure*}

\subsection{Local Energy Behavior}

\noindent We now find the local energy behavior in the vicinity of the quantum caustics, and where its nonclassical features are most pronounced.  For $\alpha = 1$, with $\hat H = -\partial_\chi^2 + \chi$ in our choice of units, the local energy function of $\psi_N(\chi,\tau)$, as given by Eqs.~(\ref{Eq-WV}), (\ref{Eq-Gauss_Cn}), and (\ref{Eq-phi_n}), is 
\begin{widetext}
    \begin{align}
    {\tilde E}_N(\chi,\tau) = -\frac{\sum_{n=1}^{N}z_n\exp\left[z_n\sigma^2+iz_n\tau\right]\Ai\left(\sigma^4+\chi_0+z_n\right)\Ai(\chi+z_n)/|\Ai'(z_n)|^2}{\sum_{n=1}^N\exp\left[z_n\sigma^2+iz_n\tau\right]\Ai\left(\sigma^4+\chi_0+z_n\right)\Ai(\chi+z_n)/|\Ai'(z_n)|^2}.
\end{align}
In $N\rightarrow\infty$ limit $\psi_N(\chi,0)\rightarrow\psi_0(\chi)$, $\tilde E_N(\chi,0)$ approaches the function
\begin{equation}
    \tilde E(\chi,0)=\lim_{N\rightarrow\infty}\tilde E_N(\chi,0) = -\frac{(\chi-\chi_0)^2}{4\sigma^4}+\frac{1}{2\sigma^2}+\chi,
\end{equation}
that is positive only in a small interval
    \begin{align}
\label{Eq-normal_Interval}
    \chi_0+ 2\sigma^4 - 2\sqrt{\sigma^8 + \chi_0\sigma^4 + \frac{1}{2}\sigma^2} <\chi<\chi_0 + 2\sigma^4+ 2\sqrt{\sigma^8 + \chi_0\sigma^4 + \frac{1}{2}\sigma^4},
\end{align}
\end{widetext}
labeled $\ell$ in Fig.~\ref{fig-Init_Ew}.
Outside of the interval (\ref{Eq-normal_Interval}), the wavefunction $\psi_0(\chi)$ superbehaves since $\tilde E(\chi,0)$ is negative.
As shown in Fig. \ref{fig-Init_Ew}, the convergence is not uniform. 
The distribution $\tilde E_N(\chi,0)$ accurately approximates $\tilde E(\chi,0)$ only within the interval given by (\ref{Eq-normal_Interval}), which is consistent with the fact that all eigenenergies $\varepsilon_n$ are strictly positive.
\par
Outside of the interval (\ref{Eq-normal_Interval}), $\psi_N(\chi,0)$ has many nodes [see Figs. \ref{Fig-Initial_State}(a) and (a${}'$)]. 
Between phase singularity $\chi_{s_j}$ and $\chi_{s_{j+1}}$ ($j=1,2,\ldots$), $\tilde E_N(\chi,0)$ alternates from $-\infty$ to $+\infty$, and vice versa, resembling $\tan\chi$ and $\cot\chi$ functions, respectively, both having a single inflection point at $\chi_{i_j}$ ($j=1,2,\ldots$).
To understand the average behavior, it is necessary to eliminate the influence of singularities that render the statistical moments undefined.
Therefore, we connect all inflection points  $(\chi_{i_j},\tilde E(\chi_{i_j}))$ by a poly-line and analyze the resulting curves instead.
On the right of the interval $\ell$ (for $\chi>15$) we have found that poly-line, $(\chi_{i_j},\tilde E(\chi_{i_j}))$ oscillates with small amplitude around the mean value $\varepsilon_N=152.529$.
On the left of $\ell$ (for $\chi<15$), as a rule $\tilde E_N(\chi_{i_j},0)>\varepsilon_N$, except at the single point $\min\tilde E_N(\chi_{i_j},0)= 152.528\approx\varepsilon_N$. 
At the coordinate origin $\tilde E(\chi_{i_j}, 0)$ reaches a value
\begin{widetext}
    \begin{align}
    \tilde E_N(0,0)=\lim_{\chi\rightarrow0}\tilde E_N(\chi,0)=-\frac{\sum_{n=1}^{N}z_n\frac{(-1)^n}{|\Ai'(z_n)|}\exp\left[z_n\sigma^2\right]\Ai\left(\sigma^4+\chi_0+z_n\right)}{\sum_{n=1}^N\frac{(-1)^n}{|\Ai'(z_n)|}\exp\left[z_n\sigma^2\right]\Ai\left(\sigma^4+\chi_0+z_n\right)}=198.783,
\end{align}
\end{widetext}
that is close to the local energy $\tilde E_{k_\psi}=k_\psi^2=211.834$ predicted on the basis of the harmonic approximation (\ref{Eq-harm_approx},\ref{Eq-loc_k}). 
The deviation between these values comes from the fact that we have taken a finite value of $N=400$. 

\begin{figure*}[!tb]
    \centering
    \includegraphics[width=1\linewidth]{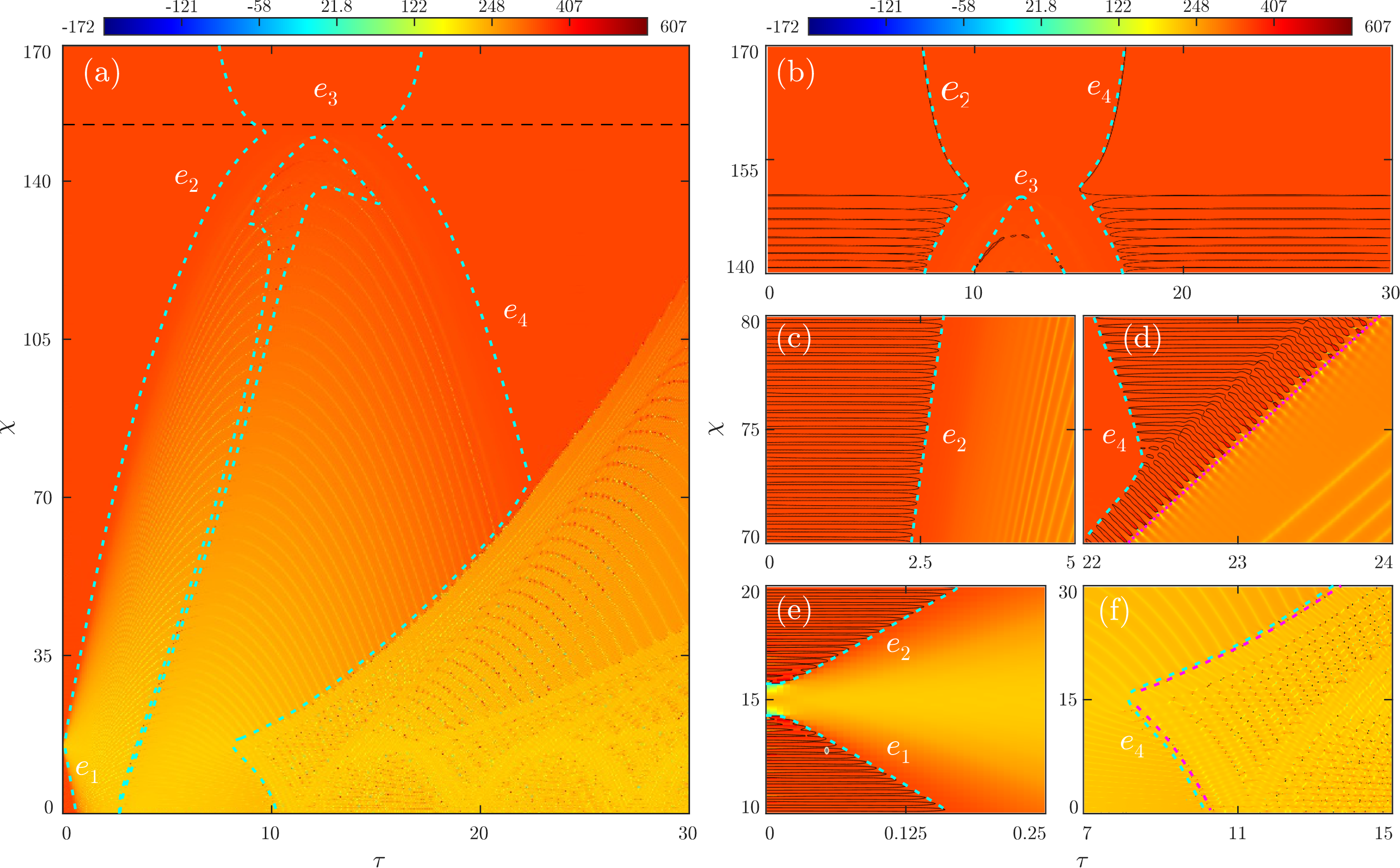}
    \caption{(a) Evolution of $\Re\{\tilde E_N(\chi,\tau)\}$ from Fig. \ref{Fig-Loc_Energy} over extended range.  
    The black dashed line marks the maximum re-bounce height $\chi=-z_N=\varepsilon_N$.
    (b-f) Enlarged views of $\Re\{\tilde E_N(\chi,\tau)\}$ in the characteristic regions.
    Thin black lines show contours $\Re\{\tilde E_N(\chi,\tau)\}=\varepsilon_N$. 
    The dashed cyan lines are encompassing curves of the contour family. 
    The magenta dashed lines indicate the dominant singularity chains from Fig. \ref{fig-Bouncing_state_zoom}  (see text for details).    
    }
    \label{Fig-Loc_Energy_Full}
\end{figure*}
\par
Figure \ref{Fig-Loc_Energy} shows evolution of $\tilde E_N(\chi,\tau)$ in region $\mathcal{D} = (0\leq\chi\leq35) \times (0\leq\tau\leq25)$.
Note that superbehaving regions around phase singularities from Fig. \ref{fig-Init_Ew} exist for some finite time before disappearing.
This occurs because the time evolution introduces energy-dependent relative phase factors which perturb interference of eigen-states $\varphi_n(\chi)$ in regions $\chi\notin \ell$.
The gradual disruption of the delicate, nearly complete destructive interference, essential for approximating $\psi_0(\chi)$, ultimately leads to the breakdown of the superbehaving structure.
In other words, the expansion of the initial wave packet, driven by wave dispersion, progressively covers the phase singularities and eventually erases them.
\par
The mean value of $\tilde E_N(\chi,\tau)$ in $\mathcal{D}$
\begin{equation}\label{Eq-mean_E_m}
    \bar E_N=\frac{1}{|\mathcal{D}|} \int\limits_\mathcal{D}\tilde E_N(\chi,\tau)\ud\chi\ud\tau=28.3737+0.1602i,
\end{equation} 
was found to be very close to the mean energy of the state $\psi_N$,
\begin{equation}
    \bar\varepsilon=\sum_{n=1}^N|c_n|^2\varepsilon_n=26.036+0i.
\end{equation}
We stress the mean value of $\tilde E_N(\chi,\tau)$ in $\mathcal{D}$ is not the expected value of the local energy in a spatial region $(0, L)$, which is given by the expectation value of the projection operator on this region times the Hamiltonian operator.  
As evident from Fig. \ref{Fig-Loc_Energy}, function $\tilde E_N(\chi,\tau)$ mostly oscillates around $\bar E_N$ with small amplitude.
The fractions of $\mathcal{D}$ where $\Re\{\tilde E_N(\chi,\tau)\}$ and $\Im\{\tilde E_N(\chi,\tau)\}$ exibit superbehavior are found to be
\begin{eqnarray}\label{Eq-super_fractions2}
&&    \delta_r=1-\frac{1}{|\mathcal{D}|}\int\limits_{\mathcal{D}}\bm1_{0\leq\Re\{\tilde E_N\}\leq\varepsilon_N}\ud\chi\ud\tau=0.035,\\
    &&
    \delta_i=1-\frac{1}{|\mathcal{D}|}\int\limits_{\mathcal{D}}\bm1_{0\leq\Im\{\tilde E_N\}\leq\varepsilon_N}\ud\chi\ud\tau=0.492,
\end{eqnarray}
in which, both quantities are predominantly negative ($\bm{1}_A(x)$ is an indicator function of set $A$).
The fractions of $\mathcal{D}$ where their magnitudes are large,
\begin{IEEEeqnarray}{rCl}\label{Eq-Super_Super_E_w}
    \bar\delta_r&=&\frac{1}{|\mathcal{D}|}\int\limits_{\mathcal{D}}\bm1_{|\Re\{\tilde E_N\}|>\varepsilon_N}\ud\chi\ud\tau=7.9\times10^{-3},\\
    \quad\bar\delta_i&=&\frac{1}{|\mathcal{D}|}\int\limits_{\mathcal{D}}\bm1_{|\Im\{\tilde E_N\}|>\varepsilon_N}\ud\chi\ud\tau=1.7\times10^{-3},
\end{IEEEeqnarray}
are confined to very small neighborhoods surrounding phase singularities present in the initial state, or to the nearly negligible regions around singularities associated with various quantum caustics.
This explains why the potential $V(\chi)$ appears to have negligible influence on the superbehavior.
\par
However, at greater heights the energy behavior changes considerably. 
To demonstrate this, we computed $\tilde E_N(\chi,\tau)$ in the larger region $\mathcal{D}' = (0\leq\chi\leq170) \times (0\leq\tau\leq30)$, extending slightly above the maximal quantum caustic height $\chi=-z_N$, and examined the family of contours $\Re\{\tilde E_N(\chi,\tau)\}=\varepsilon_N$.
If all contours were plotted in Fig. \ref{Fig-Loc_Energy_Full}(a) the resulting graph would be unintelligible, as it would be impossible to distinguish where individual contours begin and end.
Instead, Fig. \ref{Fig-Loc_Energy_Full}(a) shows a set of lines, labeled $e_1,\ldots e_4$, that encompass corresponding contours $\Re\{\tilde E_N(\chi,\tau)\}=\varepsilon_N$, thereby partitioning $\mathcal{D}'$ into regions that either contain phase singularities or are free from them.
The contour lines are given in Figs. \ref{Fig-Loc_Energy_Full}(b-f) which present enlarged views of $\Re\{\tilde E_N(\chi,\tau)\}=\varepsilon_N$ and illustrate the characteristic features of the contour distribution.
\par
The line $e_1$, [see Fig. \ref{Fig-Loc_Energy_Full}(a) and (e)], traces the lower bound of expanding main lobe of $\psi_N(\chi,0)$ (originating at $\chi_0-|\ell|/2$), and terminates at the boundary $\chi=0$.
Similarly, line $e_2$ (originating at $\chi_0+|\ell|/2$) traces the upper bound and reaches $\chi=-z_N$, along a downward curved trajectory, consistent with the deaccelerating effect of the potential $V(\chi)$ [see Fig. \ref{Fig-Loc_Energy_Full}(a-c) and (e)].
The interference of the purely evanescent waves given by Eq. (\ref{Eq-Evanescent_waves}) produces the largest contour $\Re\{\tilde E_N(\chi,\tau)\}=\varepsilon_N$ encompassed by the line $e_2$.
The line $e_3$ surround contours $\Re\{\tilde E_N(\chi,\tau)\}=\varepsilon_N$  arising from the self-interference between the expanding and reflected parts of the wave packet. 
For $0\leq\chi\leq30$ line $e_4$ nearly coincides with the singularity chains on the dark side of caustics shown in Figs. \ref{fig-Bouncing_state_zoom}(a) and (b).
The contours it encloses are barely visible in Fig. \ref{Fig-Loc_Energy_Full}(f).
However, as quantum caustic carry the probability density in the the regions of high potential energy, these contours extend perpendicularly to the singularity chain, and for $\chi>74$ also grow parallel to the $\tau$ axis [see Figs. \ref{Fig-Loc_Energy_Full}(b) and (d)].
Meanwhile, contours $\Re\{\tilde E_N(\chi,\tau)\}=\varepsilon_N$ on the bright side of the caustics $c_1$ (and all other) remain small.
The largest superbehaving contour is again found in the region $\chi>-z_N$ [see Fig. \ref{Fig-Loc_Energy_Full}(b)].
The fraction of $\mathcal{D}'$ where $|\Re\{\tilde E_N(\chi,\tau)\}|>\varepsilon_N$ is now significantly larger and equal to
\begin{equation}\label{Eq-super_E_w_full}
    \bar\delta_r=\frac{1}{|\mathcal{D}'|}\int\limits_{\mathcal{D}'}\bm1_{|\Re\{\tilde E_N\}|>\varepsilon_N}\ud\chi\ud\tau=0.251.
\end{equation}
\par
\begin{figure*}[!tb]
    \centering
    \includegraphics[width=1\linewidth]{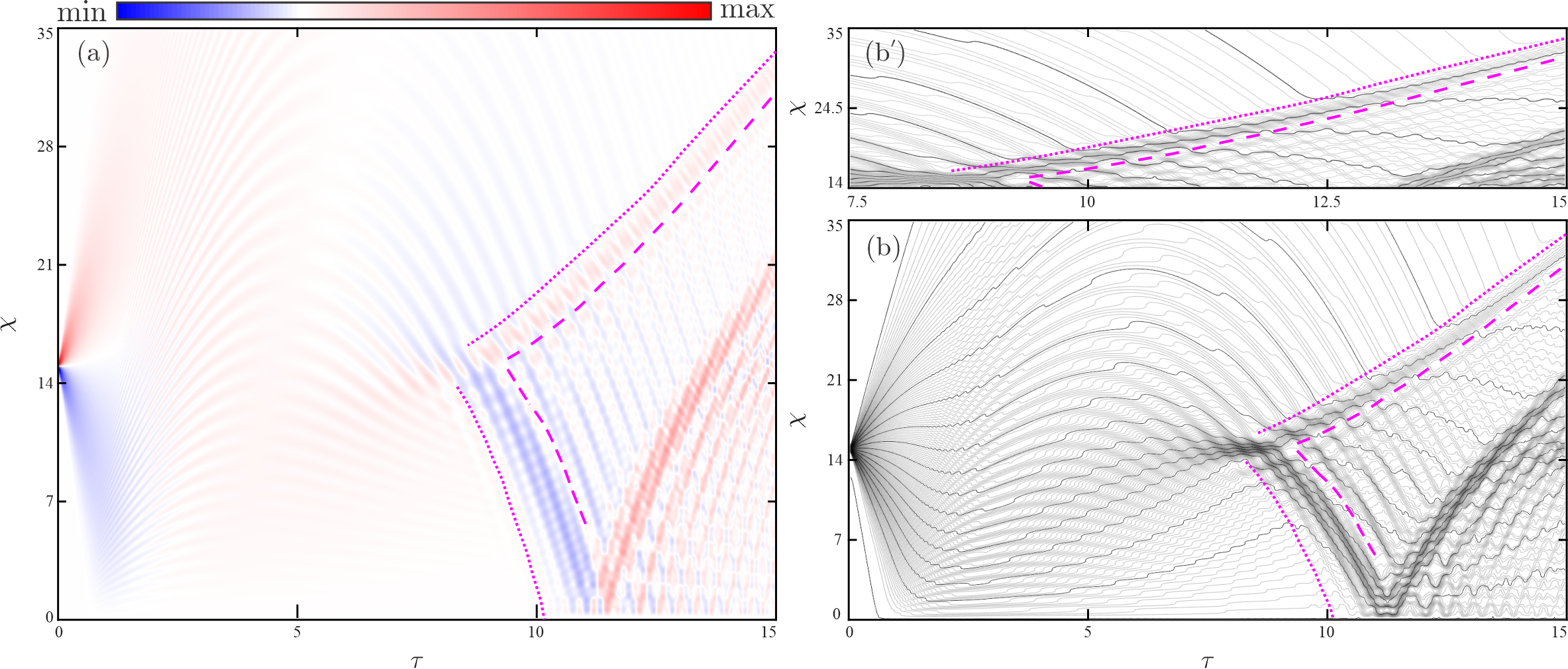}
    \caption{(a) Evolution of the probability current $J(\chi,\tau)$ for $\chi_0 = 15$, $\sigma = 0.15$, and  $N = 400$.
    (b,b') Madelung--Bohm trajectories (gray lines) for 150 initial positions sampled from $12.5\le\bar\chi_0\le17.5$. 
    For clarity, every tenth trajectory is highlighted in black. 
    The magenta dotted and dashed lines indicate the dominant and subdominant singularity chains, respectively.   Panel (b') show a zoom-in on the upward growing caustic to highlight the navigation of the Madelung-Bohm trajectories through the singularity alley.   The trajectories enter the alley between two of the dominant singularities, move {\it up} for some time, then exit between two of the subdominant singularities.}
    \label{Fig-Bohm_trajs}
\end{figure*}

\section{Madelung-Bohm trajectories and superbehavior}\label{sec:MBtraj}
\noindent In this section, we reanalyze the interference patterns that arise in the quantum bouncer from the perspective of the hydrodynamic point of view of Madelung \cite{madelung1927quantum} - later rediscovered and given an ontological interpretation by David Bohm \cite{Bohm1952I,Bohm1952II}.

As shown in Figs.~\ref{fig-Bouncing_state} and \ref{fig-Bouncing_state_zoom}, and explained in the text, both classical and quantum caustics extend into the region of very high potential energy.  
However, a physical picture of how this happens is very different.
In the classical description, the state $\psi_N$ represents a statistical ensemble of classical particles with the phase space density $W(\chi,\dot\chi,\tau)$ satisfying Liouville's equation
\begin{equation}
    \!\!\!\dot W(\chi,\dot\chi,\tau)+\dot\chi\partial_\chi\!W(\chi,\dot\chi,\tau)-V'(\chi)\partial_{\dot\chi}\!W(\chi,\dot\chi,\tau)=0.
\end{equation}
The initial distribution, consistent with the quantum description, is
\begin{equation}\label{Eq-Ph_SP_Dist}
    W(\chi,\dot\chi,0) = \frac{1}{2\pi}\exp\left[-\frac{(\chi-\chi_0)^2}{2\sigma_\chi^2}-\frac{\dot\chi^2}{2\sigma_{\dot\chi}^2}\right],
\end{equation}
where $\sigma_{\chi}=\sigma$ and $\sigma_{\dot\chi}=1/\sigma$ is deduced from the uncertainty relation $\sigma_\chi\sigma_{\dot\chi}=1$.
Although the span of the distribution (\ref{Eq-Ph_SP_Dist}) is the whole phase space, the finite sensitivity of the detector determines the minimal probability density $W_\textrm{min}$ that can be detected reliably. 
The contour $W(\chi,\dot\chi)=W_\textrm{min}$ is an ellipse whose major and minor semi-axes are $\sqrt{-2\log(2\pi W_\textrm{min})}\sigma_{\chi}$ and $\sqrt{-2\log(2\pi W_\textrm{min})}\sigma_{\dot\chi}$, respectively.
Since $\sigma_\chi\ll 1\ll\sigma_{\dot\chi}$ this ensemble can be approximated by a family of periodic parabolas (\ref{Eq-x(t)}), each starting at $\chi_0$, with the initial velocities restricted to $-\sqrt{-2\log(2\pi W_\textrm{min})}\sigma_{\dot\chi}<\dot\chi_0<\sqrt{-2\log(2\pi W_\textrm{min})}\sigma_{\dot\chi}$, whose envelopes are classical caustics.
The maximal height of the caustics
\begin{equation}
    \chi_\textrm{m}=\chi_0+\log(2\pi W_\textrm{min})\frac{2}{\sigma^2},
\end{equation}
is deduced from the conservation of energy.
Since caustics are singularities of the differential cross-section, both the spatial probability density and the current
\begin{IEEEeqnarray}{rCl}
  \rho_\textrm{c}(\chi,\tau)&=&\int\limits_{-\infty}^\infty W(\chi,\dot\chi,\tau)\ud\dot\chi,  \\ J_\textrm{c}(\chi,\tau)
&=&\int\limits_{-\infty}^\infty\dot\chi W(\chi,\dot\chi,\tau)\ud\dot\chi,
\end{IEEEeqnarray}
diverge on caustics.
However, classical caustics are limiting curves that tangentially touch each trajectory. 
This implies that no trajectory follows the caustics.
\par 
We can write any single particle state in its polar form,
\be
\psi(\chi, \tau) = R(\chi, \tau) e^{i \phi(\chi, \tau)},
\ee
where $R$ is the amplitude and $\phi$ is the phase, and both functions depend on position and time.
In the quantum description of the problem, the state $\psi_N(\chi,\tau)$ is a finite superposition of states of definite energy (\ref{Eq-Psi_x_t}) whose evolution describes, a flow of fictitious \emph{probability fluid} via continuity equation
\begin{equation}
    \partial_\tau\rho(\chi,\tau)+\partial_\chi J(\chi,\tau)=0.
\end{equation}
Its streamlines $\bar\chi(\tau;\bar\chi_0)$, known as Madelung-Bohm trajectories \cite{madelung1927quantum}, satisfy the following differential equation
\begin{equation}\label{Eq-MB-trajs}
    \dot{\bar\chi}(\tau;\bar\chi_0)=\frac{J(\bar\chi(\tau),\tau)}{\rho(\bar\chi(\tau;\bar\chi_0),\tau)}=2\phi'(\chi,\tau)|_{\chi=\bar\chi(\tau;\bar\chi_0)},
\end{equation}
where $\bar\chi(0;\bar\chi_0)=\bar\chi_0$.
and can be interpreted as conditional averaged paths of particles, operationally given by weak measurements of momentum followed postselection on position \cite{Shalm2010,Kocsis2011,Mahler2016}, that in the limit of very weak measurement, coincide with the trajectory family $\bar\chi(\tau;\bar\chi_0)$ \cite{Wiseman_2007}.
The Bohmian picture is that the guidance equation (\ref{Eq-MB-trajs}) is the same as 
\be
v_{\textrm{MB}} = 2\phi'.
\ee
where $v_\textrm{MB}$ is the Madelung-Bohm particle velocity.
Extensions to multiple dimensions are straightforward. 
The particle current $J$ is simply given by $J = v_\textrm{MB} \rho$, where $\rho = R^2$ is the local particle density.
The Schr\"odinger equation is then equivalent to two differential equations for $R$, $\phi$ (corresponding to the real and imaginary part), the first of which (analogous to the Hamilton-Jacobi equation) is
\be
{\dot \phi} = - (\phi')^2  - V_\textrm{eff}, \qquad V_\textrm{eff}(\chi) = V(\chi)-\frac{R''}{R},
\ee
\noindent where $V_\textrm{eff}$ is the so-called \emph{effective quantum potential} \cite{Bohm1952I,Bohm1952II}, containing the physical potential $V$ and a term involving the particle density and its derivatives, $V_Q=-R''/R$ called quantum potential. 
The second equation is given by
\be
{\dot R} = -\frac{1}{R}(R^2 \phi')', 
\ee
which is equivalent to the continuity equation $\partial_\tau \rho +  \partial_\chi J=0$.
It is a natural question what, if any, meaning the local energy ${\tilde E}(\chi)$ has in the Madelung-Bohm picture (\ref{Eq-WV}).  
We can calculate this quantity as before in two different ways - the first involving the spatial Hamiltonian operator $H = -\partial_\chi^2 + V(\chi)$ and the second using the time operator $H = i \partial_\tau$.
We find the real part has the form
\be
\Re\{\tilde E\} = \frac{1}{4} v_\textrm{MB}^2 + V_\textrm{eff} = - {\dot \phi},
\ee
that is, simply the sum between the kinetic energy and the quantum potential (the extra factor of 2 comes from our choice of units), while the imaginary part is given by
\be
\Im\{\tilde E\} = - \frac{J'}{2 \rho} = \frac{{\dot \rho}}{2 \rho},
\ee
or the divergence of the local current, divided by the density.
While the real part of the local energy has a clear interpretation, further insight into the physical significance of the imaginary part of weak values is given in Ref.~\cite{Dressel2012ImaginaryWV} - the imaginary part of the local energy represents the logarithmic directional derivative of the particle density along the flow of the Hamiltonian.
Thus, the local energy has a natural interpretation within the framework of the Madelung-Bohm hydrodynamic picture.
\par
We now apply these insights to the problem of quantum caustics in the bouncer problem.  Quantum caustics are the semiclassical images of the classical singularities of  $\rho_\textrm{c}$ and $J_\textrm{c}$, and their maximal extent $-z_N$ is determined by the turning point of the highest allowed excited state.
The quantum probability current $J(\chi,\tau)$, shown in Fig. \ref{Fig-Bohm_trajs}(a) remains finite at caustics because Madelung-Bohm trajectories, shown in Fig. \ref{Fig-Bohm_trajs}(b), cannot cross \cite{Bohm1952I,Bohm1952II,Holland1993}.
Instead, they cluster in high-density regions, producing apparent darkening near caustics due to the lightness-assimilation optical illusion \cite{devalois1988spatial}. 
Many trajectories temporarily run parallel to caustics before detaching [see highlighted trajectories in Fig. \ref{Fig-Bohm_trajs}(b) and (b${}'$)]. 
This classically forbidden mode of motion is related to the superbehavior as we shall now explain.
\par
Since the probability density vanishes at a wave node, no trajectory can pass through it.
When encountering a phase singularity, Madelung/Bohm trajectories must circle around it \cite{Berry1974Nye,Holland1993,Waegell2024a}.
The divergence $\Re\{\tilde E_N(\chi,\tau)\}\rightarrow\infty$ at the singularity compensates the opposite divergence of the quantum potential \cite{Holland1993}
\begin{equation}
    V_Q(\chi,\tau) = -\frac{R''(\chi,\tau)}{R(\chi,\tau)}\rightarrow -\infty,
\end{equation}
and explain how it allows particles to avoid being trapped by the infinitely deep local minimum of the quantum potential \cite{Waegell2024a}. 
As trajectories approach chains of the dominant singularities, the described circulation transfers them from the dark to the bright side of the caustic.
They then move within the region bounded between dominant and subdominant chains of singularities, we call the ``singularity alley''.
As a result, circulation around a dominant singularity bends the trajectory and will then trap the Madelung/Bohm trajectories within the singularity alley - parallel to the classical caustic! - for some time before they are finally ejected out the other side, around a subdominant singularity. 
The same mechanism applies to other caustic crossings.

\section{Conclusions} \label{sec:conc}\noindent
We have investigated the caustic formation in the quantum bouncing ball problem, starting from a Gaussian initial state.  Surprisingly rich structures appear in the resulting waveforms of the ball as it bounces up and down. The wave patterns can be understood as a superposition of the canonical cusp diffraction catastrophes.
One of our findings is that the quantum caustics are built on the `scaffold' of the saddle-node bifurcation of the classical caustic formation.  We have classified this as a cusp-type catastrophe, and modeled it with an Arnol'd theory universal polynomial prototype.  Indeed, we find that modeling the quantum caustic with a Pearcey function with an Arnol'd action reproduces quite well the basic features of the numerically found interference pattern.
In particular, the resulting orderly pattern of wave nodes reflects the corresponding distribution of phase singularities of the individual Pearcey function.
While we have illustrated this result with the bouncing ball problem, we stress the generic nature of this effect. 
According to the Whitney theorem \cite{whitney1955singularities}, each structurally stable distribution of caustics in 2D can be modeled as a collection of   whose semiclassical model is given by a combination of Airy and Pearcey functions \cite{Arnold1972IntegralsOR,PostonStewart1996Catastrophe}
The dominant phase singularities lie on the dark side of the caustics, which carry them into the regions of high potential energy.
It is shown that a large interaction potential $V(\chi)$ increases the size of the superbehaving regions that surround wave nodes.
\par
While the Gaussian wavepacket extends to arbitrary large energies, any finite truncation of this waveform on an energy limited set of energy eigenfunctions can lead to superenergy behavior.  We showed numerically that the local energy can exceed the highest energy in the superposition at numerous places in space and time - notably near surrounding phase singularities - and quantified their fraction.  This effect is magnified when the spatial region extends above the highest quantum caustic.

\par
We gave a complimentary analysis of this effect using a quantum hydrodynamic description with a Madelung-Bohm approach. We find that the density of Madelung-Bohm trajectories is high near caustics, and the trajectories undergo a rapid wiggling as they traverse the phase singularity chains, bending to move along the classical caustic line, in order to avoid the phase singularities as they first pass the dominant singularity chain, are trapped in the alley, and eventually escape through the subdominant singularity chain.
One distinguishing feature of the Madelung-Bohm trajectories is that by definition, only one point on each classical trajectory belongs to the caustics, while in fact many Madelung-Bohm trajectories run parallel to the caustics for finite periods.

\section{Acknowledgments}

\noindent
ANJ and M\'C thank Dr. Mordecai Waegell for his helpful comments on the paper and valuable discussion of the results.
M\'C acknowledges the Institute for Quantum Studies at Chapman University for providing the facilities and support during a scientific visit in academic year 2025/2026, realized with the support of the Fulbright Scholar program No. G-1-0005 and the Ministry of Science, Technological Development, and Innovation of the Republic of Serbia through grant No. 451-03-33/2026-03/ 200017.  ANJ acknowledges support through the Air Force Office of Scientific Research under award number FA9550-24-1-0329 and the John Templeton Foundation, under grant ID 63209.

\appendix
\section{Fourier-Airy expansion of Gaussian function\label{App-AF_Gauss}}
\noindent The coefficients of Fourier-Airy \cite{vallee2010airy} expansion of Gaussian (\ref{Eq-Psi_0}) are 
\begin{equation}\label{Eq-F-Ai_Gaussian}
    c_n = \frac{\frac{\alpha^{1/6}}{\sqrt[4]{2\pi\sigma^2}}}{|\Ai'(z_n)|}\int\limits_0^\infty\Ai(\alpha^{1/3}\chi+z_n)e^{-\frac{(\chi-\chi_0)^2}{4\sigma^2}}\ud\chi.
\end{equation}
For $\sigma\ll 1$, the lower bound of integration can be safely extended towards $-\infty$. 
Using the integral representation of the Airy function 
\begin{equation}\label{App-Eq-Airy_rep}
    \Ai(z)=\frac{1}{2\pi}\int\limits_{-\infty}^\infty\exp\left[i\frac{1}{3}\eta^3+iz\eta\right]\ud\eta.
\end{equation}
Eq. (\ref{Eq-F-Ai_Gaussian}) becomes
\begin{widetext}
\begin{align}
    c_n = \frac{1}{2\pi}\frac{1}{\sqrt[4]{2\pi\sigma^2}}\frac{\alpha^{1/6}}{|\Ai'(z_n)|}\int\limits_{-\infty}^\infty\int\limits_{-\infty}^\infty\exp\left[i\frac{1}{3}\eta^3-\frac{(\chi-\chi_0)^2}{4\sigma^2}+i(\alpha^{1/3}\chi+z_n)\eta\right]\ud\chi\ud\eta.
\end{align}
The identity \cite{AbramowitzStegun1972Handbook}
\begin{equation}\label{App-Eq-Airy-Identity}
    \int\limits_{-\infty}^\infty\exp[-az^2+2bz+c]\ud z = \sqrt{\frac{\pi}{a}}\exp\left[-\frac{b^2}{a}-c\right],
\end{equation}
allows integral over $\chi$ to be evaluate analytically giving
\begin{align}
    c_n = \frac{2^{3/4}\pi^{1/4}\sigma^{1/2}}{2\pi}\frac{\alpha^{1/6}}{|\Ai'(z_n)|}\int\limits_{-\infty}^\infty\exp\left[i\frac{1}{3}\eta^3-\alpha^{2/3}\sigma^2\eta^2+i(\alpha^{1/3}\chi_0+z_n)\eta)\right]\ud\eta.
\end{align}
Applying the integral identity \cite{vallee2010airy} 
\begin{equation}
    \int\limits_{-\infty}^\infty\exp[\frac{i}{3}\eta^3+ia\eta^2+ib\eta]\ud\eta=2\pi\exp\left[ia\left(2a^2/3-b\right)\right]\Ai(b-a^2),
\end{equation}
with $\alpha=1$ we get relation (\ref{Eq-C_n}) given in the text.

\section{The small-\texorpdfstring{$\chi$}{chi} asymptotics\label{App-small_chi_asympt}}
\noindent
For $\alpha=1$, the approximation of the initial state $\psi_0$ is
\begin{equation}
\psi_N(\chi) = \sum_{n=1}^N \frac{(8\pi\sigma^2)^{1/4}}{(\Ai'(z_n))^2}
\exp\!\left(\frac{2}{3}\sigma^6 + (\chi_0 + z_n)\sigma^2\right)
\mathrm{Ai}(\sigma^4 + \chi_0 + z_n) \Ai(\chi + z_n).
\end{equation}
We are looking for the asymptotic form of $\psi_N(\chi)$ for $0 < \chi < |z_1|$.
In that case, all arguments of Airy functions are negative so $\Ai(\eta)$ and $\Ai'(\eta)$ can be approximated by the $-|\eta|\gg 1$ formulas
\begin{equation}
\Ai(\eta) \approx \frac{1}{\sqrt{\pi}|\eta|^{1/4}}
\sin\!\left(\frac{2}{3}|\eta|^{3/2} + \frac{\pi}{4}\right),\quad 
\Ai'(\eta) \approx -\frac{|\eta|^{1/4}}{\sqrt{\pi}} 
\cos\!\left(\frac{2}{3}|\eta|^{3/2} + \frac{\pi}{4}\right),
\end{equation}
and $\Ai$ zeros by $n\gg1$ formula $z_n\approx-[3\pi/2(n-1/4)]^{2/3}$.
Consequently
\begin{equation}
\frac{2}{3}|z_n|^{3/2} + \frac{\pi}{4} = n\pi,\Rightarrow |\mathrm{Ai}'(z_n)|^2 \sim \frac{|z_n|^{1/2}}{\pi}.
\end{equation}
Using binomial expansions
\begin{equation}
(|z_n| - \chi)^{3/2}
\approx |z_n|^{3/2} - \frac{3}{2}|z_n|^{1/2}\chi,\quad\textrm{and}\quad (|z_n| - \chi)^{1/4}\approx |z_n|^{1/4},
\end{equation}
on gets
\begin{align}
\Ai(\chi + z_n)\approx\frac{1}{\sqrt{\pi}|z_n|^{1/4}}\sin\left(n\pi - |z_n|^{1/2}\chi \right) = \frac{(-1)^{n+1}}{\sqrt{\pi}|z_n|^{1/4}} \sin\!\left(|z_n|^{1/2}\chi\right).
\end{align}
Thus, close to coordinate origin $\psi_N(\chi)$ is given by following Generalized Fourier sequence \cite{Aharonov2017Mathematics}
\begin{equation}
\psi_N(\chi)\approx(8\pi^3\sigma^2)^{1/4} \sum_{n=1}^N(-1)^{n+1}\frac{\exp\left[\frac{2}{3}\sigma^6+(\chi_0 +z_n)\sigma^2\right]}{|z_n|^{3/4}}\mathrm{Ai}(\sigma^4+\chi_0 + z_n)\sin\left(|z_n|^{1/2}\chi\right),
\end{equation}
that for $\sigma^2\ll1$ further simplifies to 
\begin{equation}
    \psi_N(\chi)\approx(8\pi^3\sigma^2)^{1/4} \sum_{n=1}^N(-1)^{n+1}\frac{\exp\left[(\chi_0 +z_n)\sigma^2\right]}{|z_n|^{3/4}}\mathrm{Ai}(\chi_0 + z_n)\sin\left(|z_n|^{1/2}\chi\right).
\end{equation}
Obtained sine-series coincides with Eq. (\ref{Eq-Fur_Sec-approx}) in the main text, whose amplitudes are given by Eq. (\ref{Eq-Exp_coeff}) of the main text.
\clearpage
\end{widetext}

\section{Catastrophic modeling of the quantum caustics\label{App-catatrophe-fit}}
\noindent
The graph of the canonical semicubical parabola consists of two branches
\begin{equation}\label{App-Eq-Sem_cub_parab}
    y = \pm \sqrt{-8x^3/27},
\end{equation}
both extending into $x\leq 0$ part of the coordinate plane from the cusp point at the origin. 
Note that the semicubical parabola (\ref{App-Eq-Sem_cub_parab}) is self-similar to the scaling transformation $x\rightarrow\sqrt[3]{\kappa}x$ and $y\rightarrow\sqrt{\kappa}y$, with the scale factor $\kappa>0$.
Thus, the most general linear model of the diffraction catastrophe in the $(\tau,\chi)$ space with cusp at $C=(\tau_C,\chi_C)$ and branches extending into the region $\tau\ge\tau_C$ is generated by the following polynomial
\begin{equation}
    A_3(\eta) = \eta^4+\sqrt[3]{\kappa}\frac{\tau_C-\tau}{\bar\tau_C}\eta^2+\sqrt{\kappa}\frac{\chi-\chi_C}{\bar\chi_C}\eta,
\end{equation}
whose bifurcation set
\begin{equation}
    27\left(\frac{\chi-\chi_{C}}{\bar\chi_{C}}\right)^2+8\left(\frac{\tau_{C}-\tau}{\bar\tau_{C}}\right)^3=0,
\end{equation}
does not depend on the scaling parameter $\kappa$.
The ratio $\bar\chi/\tau_c^{3/2}$ determines the inclination of the biffuctaion line at any point $\tau>\tau_C$, and can be used to force the bifurcation line to pass through the arbitrary point of the plane $C_1 = (\tau_{C_1},\chi_{C_1})$ having $\tau_{C_1}>\tau_C$.
Application of the stationary phase method \cite{Dingle1973} to points $(\tau=\bar\tau+\delta\tau,\chi+\delta\chi)$ in the small surroundings of the point $(\bar\tau,\bar\chi)$ on the bright side of the caustics, and far from the cusp, transforms the Pearcey integral into a sum
\begin{widetext}
\begin{align}
    \textrm{P}\left(\sqrt[3]{\kappa}\frac{\tau_C-\tau}{\bar\tau_C},\sqrt{\kappa}\frac{\chi-\chi_C}{\bar\chi_C}\right)\approx\sum_{n=1}^3\sqrt{\frac{2\pi}{|\frac{\ud^2}{\ud\eta^2}A_3(\eta_n)|}}\exp\left[ iA_3(\eta_n)+i\frac{\pi}{4}\textrm{sgn}\left(\frac{\ud^2}{\ud\eta^2}A_3(\eta_n)\right) \right],
\end{align}
\end{widetext}
where $\eta_n$ are the three real solutions of the cubic equation
\begin{equation}\label{App-Eq-crit_set}
    \frac{\ud A_3(\eta_n)}{\ud\eta}= 4\eta_n^3+2\sqrt[3]{\kappa}\frac{\tau_C-\tau}{\bar\tau_C}\eta_n+\sqrt{\kappa}\frac{\chi-\chi_C}{\bar\chi_C}=0.
\end{equation}
Searching for solutions of Eq.~(\ref{App-Eq-crit_set}) perturbatively in the form $\bar\eta_n+\delta\eta$ up to the first order one gets
\begin{equation}\label{App-Eq-crit_set-1}
    \begin{IEEEeqnarraybox}[][c]{c}
        4\bar\eta_n^3+2\sqrt[3]{\kappa}\frac{\tau_C-\bar\tau}{\bar\tau_C}\bar\eta_n+\sqrt{\kappa}\frac{\bar\chi-\chi_C}{\bar\chi_Cc}=0,\\
        \delta\eta=\frac{2\sqrt[3]{\kappa}
        \bar\eta_n\delta\tau/\bar\tau_C-\sqrt{\kappa}\delta\chi/\bar\chi_C}{12\bar\eta_n^2+2\sqrt[3]{\kappa}(\tau_C-\bar\tau)/\bar\tau_C},
    \end{IEEEeqnarraybox}
\end{equation}
and consequently
\begin{equation}\label{App-Eq-crit_set-2}
    \begin{IEEEeqnarraybox}[][c]{c}
        A_3(\eta_n) = A_3(\bar\eta_n)-\sqrt[3]{\kappa}\frac{\bar\eta_n^2}{\bar\tau_C}\delta\tau+\sqrt{\kappa}\frac{\bar\eta_n}{\bar\chi_C}\delta\chi,\\
        \frac{d^2}{d\eta^2}A_3(\eta_n)= \frac{d^2}{d\eta^2}A_3(\bar\eta_n)+24\bar\eta_n\delta\eta-2\sqrt[3]{\kappa}\frac{\delta\tau}{\bar\tau_C}.
    \end{IEEEeqnarraybox}
\end{equation}
Thus, the parameter $\kappa$ can be used to set the local wave vectors of the interfering weaves
\begin{equation}
    \bm{k}_n=\left(-\sqrt[3]{\kappa}\frac{\bar\eta_n^2}{\bar\tau_c},\sqrt{\kappa}\frac{\bar\eta_n}{\bar\chi_c}\right),
\end{equation}
and thereby influence the positions of its phase singularities.
\par
The catastrophic model of the semiclassical caustics $C_1$ from Fig.~\ref{fig-Bouncing_state}(a) was determined so that the cusp of its bifurcation set is at $C_1=(\tau_{C_1},\chi_{C_1})$. 
This sets $(\tau_C,\chi_C)=(\tau_{C_1},\chi_{C_1})$.
Additionally, the lower branch of the bifurcation was required to pass through the point $D_1=(\tau_{D_1},\chi_{D_1})=(10.954,0)$ and that first phase singularity of $\textrm{P}(x,y)$ of the first singularity of the lower chain $Z=(x_z,y_z)=(-4.378,-0.527)$ \cite{BerryNyeWright1979EllipticUmbilic} coincides with the point $s_1=(\tau_{s_1},\chi_{s_1})=(9.33,14.798)$.
The described requirements are expressed in the system of equations
\begin{IEEEeqnarray}{c}\label{App-Eq-fit_sys}
    \frac{\bar\chi_C^2}{\bar\tau_C^3}=\frac{27}{8}\frac{(\chi_{D_1}-\chi_C)^2}{(\tau_{D_1}-\tau_C)^3},\\
    \sqrt[3]{\kappa}\frac{\tau_C-\tau_{s_1}}{\bar\tau_C}=x_z, \quad\sqrt{\kappa}\frac{\chi_{s_1}-\chi_C}{\bar\chi_C}=y_z,     
\end{IEEEeqnarray}
which is consistent only if
\begin{equation}\label{App-Eq-consist}
    \frac{27}{8}\frac{(\chi_{D_1}-\chi_C)^2}{(\tau_{D_1}-\tau_C)^3}=\frac{x_z^3}{y_z^2}\frac{(\chi_{S_1}-\chi_C)^2}{(\tau_{S_1}-\tau_C)^3}.
\end{equation}
If the catastrophic model is only an approximation of a more detailed model (as here), the consistency condition (\ref{App-Eq-consist}) may not hold.
In that case, the system Eq. (\ref{App-Eq-fit_sys}) is solved approximately, in the last square sense, using the freedom to choose a scaling parameter $\kappa$ to minimize the cost function
\begin{equation}
    \!\!f(\kappa) = \left[\kappa-\frac{x_z^3\bar\tau_C^3}{(\tau_C-\tau_{S_1})^3}\right]^2\!+\left[\kappa-\frac{y_z^2\bar\chi_C^2}{(\chi_{S_1}-\chi_C)^2}\right]^2\!\!.
\end{equation}
\par 
If we set $\kappa=1$, then the solution of the system (\ref{App-Eq-fit_sys})
\begin{IEEEeqnarray}{rCl}   
        \bar\tau_C &=& \frac{8}{27}\frac{x_z^2}{y_z^2}\frac{(\chi_{s_1}-\chi_C)^2(\tau_{D_1}-\tau_C)^3}{(\chi_{D_1}-\chi_C)^2(\tau_C-\tau_{s_1})^2}=0.048,\\
        \bar\chi_C &=& \frac{x_z}{y_z}\frac{\chi_{s_1}-\chi_C}{\tau_C-\tau_{s_1}}\bar\tau_C=0.052,    
\end{IEEEeqnarray}
guarantees that the bifurcation set passes the point $D_1$.
Consequently, the optimal value of the parameter $\kappa$ is given by $\partial_\kappa f(\kappa)=0$, implying
\begin{equation}
     \kappa = \frac{1}{2}\frac{x_z^3\bar\tau_C^3}{(\tau_C-\tau_{S_1})^3}+\frac{1}{2}\frac{y_z^2\bar\chi_C^2}{(\chi_{S_1}-\chi_C)^2}=0.010,
\end{equation}
and the resulting catastrophic model of the wave function near $C_1$ from Fig. \ref{fig-Bouncing_state_zoom}(b) is
\be
\begin{IEEEeqnarraybox}[][c]{rCl}
    \psi_N(\chi,\tau)&\approx&\textrm{P}\left(\sqrt[3]{\kappa}\frac{\tau_C-\tau}{\bar\tau_C},\sqrt{\kappa}\frac{\chi-\chi_C}{\bar\chi_C}\right)\\
    &=&\textrm{P}\left(\frac{2\sqrt{15}-\tau}{0.224},\frac{\chi-15}{0.509}\right).
\end{IEEEeqnarraybox}
\ee

\section{Numerical Implementation\label{App-Numerics}}
\noindent Wave functions in Figs.~\ref{Fig-Initial_State}–\ref{fig-Bouncing_state_zoom} were computed by evaluating the analytical expressions on a uniform rectangular grid with 2001 spatial and 5001 temporal points.
Airy function and its derivatives were evaluated using a specialized Bessel-function library \cite{Amos1986Algorithm644}, and the Pearcey function using the cuspint library \cite{Kirk2000Cuspint}.
The same grid was used to compute distributions $\tilde E(\chi,\tau)$ and $J(\chi,\tau)$ shown in Figs.~\ref{fig-Init_Ew}, \ref{Fig-Loc_Energy} and \ref{Fig-Bohm_trajs}(a), while a larger $5001\times5001$ grid was used to evaluate the distribution shown in Fig.~\ref{Fig-Loc_Energy_Full}.
\par
The contours $\Re\{\tilde E(\chi,\tau)\}=\varepsilon_N$ shown in Fig. \ref{Fig-Loc_Energy_Full} were evaluated using Marching cube algorithm \cite{Lorensen1987MarchingCubes}.
Integrals (\ref{Eq-mean_E_m}), (\ref{Eq-Super_Super_E_w}), and (\ref{Eq-super_E_w_full}) were evaluated using the trapezoidal rule \cite{AbramowitzStegun1972Handbook}.
Instead of solving differential Eq. (\ref{Eq-MB-trajs}), we have relied on the fact that 1D Madelung-Bohm trajectories have the property \cite{Brandt1998QuantileMotion}
\begin{equation}    \int\limits_{\bar\chi(0;\bar\chi_0)}^{\bar\chi(\tau;\bar\chi_0)}\rho(\chi,\tau)\ud\chi = \textrm{const}.
\end{equation}
The quantum trajectories, shown in Fig.~\ref{Fig-Bohm_trajs}, are quantile contours (obtained by the Marching Cube algorithm) of the cumulative probability density, computed numerically using the trapezoidal rule.

\bibliography{refs}

\end{document}